# Towards Daily High-resolution Inundation Observations using Deep Learning and EO

Antara Dasgupta[1,2,*], Lasse Hybbeneth[1], Björn Waske[1]

[1]Remote Sensing Working Group, Institute of Informatics, University of Osnabrück, Germany (antara.dasgupta@uos.de)

[2]Water Group, Department of Civil Engineering, Monash University, Australia

*Corresponding Author

## Abstract

Large-scale flood inundation observations, not only provide crucial information for emergency response and decision-making but also for future preparedness or damage assessment. Satellite remote sensing presents a cost-effective solution for synoptic flood monitoring, and satellite-derived flood maps provide a computationally efficient alternative to numerical flood inundation models traditionally used. While satellites do offer timely inundation information when they happen to cover an ongoing flood event, they are limited by their spatiotemporal resolution in terms of their ability to dynamically monitor flood evolution at various scales. Constantly improving access to new satellite data sources as well as big data processing capabilities has unlocked an unprecedented number of possibilities in terms of data-driven solutions to this problem, specifically using deep-learning algorithms with multi-sensor remotely sensed data. Specifically, the fusion of data from satellites, such as the Copernicus Sentinels, which have high spatial and low temporal resolution, with data from NASA's SMAP and GPM missions, which have low spatial but high temporal resolutions could yield high-resolution flood inundation at a daily scale. Here a Convolutional-Neural-Network is trained using flood inundation maps derived from Sentinel-1 Synthetic Aperture Radar and various hydrological, topographical, and land-use based predictors for the first time, to predict high-resolution probabilistic maps of flood inundation. The performance of UNet and SegNet model architectures for this task is evaluated, using flood masks derived from Sentinel-1 and Sentinel-2, separately with 95%-confidence intervals. The Area under the Curve (AUC) of the Precision Recall Curve (PR-AUC) is used as the main evaluation metric, due to the inherently imbalanced nature of classes in a binary flood mapping problem, with the best model delivering a PR-AUC of ~0.85. Feature importance analysis using the permutation feature importance method, showed low importance of precipitation and soil-moisture, due to the large spatial resolution mismatch and the consequent lack of spatial variability per output pixel. Results from this proof-of-concept study indicate that multi-sensor data fusion could yield daily scale high-resolution flood inundation maps, enabling new possibilities for flash flood and dynamic flood evolution monitoring using satellites.

# 1 Introduction

The Emergency Event Database (EM-DAT) records from 1970-2019 showcase the unsettling reality that weather, climate and water hazards occurred daily on average during this period, resulting in 115 fatalities and US$202M in losses every day, with average reported losses growing seven-fold from 1970–1979 to 2010–2019 (World Meteorological Organization, 2019). As population exposure increases under future climates, the impact of such hazards are only expected to intensify (Mitchell et al., 2022). There is, thus, an urgent need to develop collaborative approaches to monitor, forecast, and prepare for such events on a global scale (IPCC, 2022).

A common feature of weather, climate and water hazards is the associated inundation, typically occurring as flooding over usually dry land (Dasgupta et al., 2018). Accurate and representative estimates of inundation extent and its spatiotemporal dynamics are therefore fundamental to understand and manage flood risk as well as water resources (Fleischmann et al., 2022). Satellites provide cost-effective monitoring solutions for inundation due to their synoptic views and global coverage, and their ability to objectively observe with implicit considerations of changes in climate, land use and infrastructure (Tellman et al., 2021). Indeed, uncertainties in global flood models lead to huge discrepancies in inundation estimates (Trigg et al., 2016), lack of transferability from regional to local scales (Fleischmann et al., 2019), and over-estimation of smaller, more frequent flood events (Hawker et al., 2019), thus, necessitating the development of methods to ensure comparability (Hoch & Trigg, 2019).

Despite their growing utility and ubiquity, the challenge with satellite-derived flood extents remains in their instantaneous nature, often unable to capture flood inundation dynamics and in case of fast moving floods, they may miss the inundation completely (Tarpanelli et al., 2022). The inability to reliably capture peak inundation using satellite remote sensing, is largely due to their low systematic revisit time and unequal acquisition frequency based on regional goals (Apicella et al., 2022; Sawyer et al., 2022). Moreover, the Dartmouth Flood Observatory reports average flood duration between 1985 and 2003 as 9.5 days with a median duration of 5 days, making observing them using satellites with longer revisit times mostly a function of luck (DFO, 2004). In fact, Tarpanelli et al., 2022, show that the medium resolution public satellite imagery from the Copernicus Sentinels, catch only about 60% of flood events in Europe. Note that this analysis based on the 3-day revisit of the Sentinel-1 constellation in Europe, does not account for the failure of Sentinel-1b earlier this year or the vast discrepancies in observation frequency in the rest of the world.

Synthetic Aperture Radar (SAR) satellites currently form the benchmark in flood inundation mapping, due to their cloud and rain penetration capabilities as well as their active imaging technique, independent of solar illumination (Shen et al., 2019). Indeed the recently operational Global Flood Monitoring System (GFMS) by the Copernicus Emergency Management System, leverages the Sentinel-1 SAR satellites for detecting inundation at global scales (Salamon et al., 2021). The trade-off between different types of resolution forms a well-known problem in remote sensing (McCabe et al., 2017). While the spatial resolution and sensor characteristics of Sentinel-1 (S1) are nearly ideal for inundation mapping (Matgen et al., 2020), the temporal resolution still makes it challenging to use such data for consistent event monitoring (Bernhofen et al., 2022). Higher temporal resolution has also been identified as a need for insurance instruments (Tellman et al., 2022), as it allows to better capture the inundation dynamics which are naturally relevant to the calculation of pay-outs (Benami et al., 2021). The recent advances in the field of deep learning in Earth Observation, provide an opportunity to leverage more frequent coarse resolution datasets from multiple sensors, to generate more frequent inundation observations (Shen, 2018).

Deep learning (DL) methods in flood mapping, for example, convolutional neural networks (CNNs; LeCun et al., 2015), are perfectly poised to speed up real-time applications by producing outcomes close to physics based hydrodynamic models (Guo et al., 2021). Indeed, the spatial and temporal inductive biases in CNNs (Ma et al., 2019), allows them to learn inherent patterns and produce surrogate models (Hauswirth et al., 2022)which could increase processing speeds (over physics-based hydrodynamic models) by up to three orders of magnitude without compromising the accuracy (Bentivoglio et al., 2021). Moreover, as access to training data in flood mapping improves (Martinis et al., 2022), CNNs are also rapidly outperforming operational rule-based flood map processing chains and are almost set to replace most physics based algorithms soon (Helleis et al., 2022). Studies have also successfully demonstrated the possibility of leveraging synthetic training data also generated by DL models (Yokoya et al., 2020) or fusing optical and SAR imagery (Drakonakis et al., 2022), to mitigate the inherent uncertainties in each during training (Konapala et al., 2021). While applications of machine learning in flood mapping have primarily focussed on identifying inundation from SAR images (Jiang et al., 2021; Zhang et al., 2021), more recent advances also include the possibility of flood mapping directly on-board low-cost satellites (Mateo-Garcia et al., 2021). Despite the progress, only a few studies have attempted to predict flood inundation observations at high-spatial resolution, using temporally frequent remote sensing data and DL.

One of the first attempts in this direction, leveraged multispectral imagery from Landsat, dual-polarized SAR, and digital elevation models (DEMs) to produce flood maps at moderate (30 m) spatial resolution (Muñoz et al., 2021). With a focus on simulating compound flooding, Muñoz et al. (2021) presented a modified CNN and data fusion framework trained using hydrodynamic model results to the inundation resulting from hurricanes in the southeast Atlantic coast of USA. However, due to the model dependence on SAR imagery as an input feature, this application was limited by the revisit frequency which in the case of Hurricane Matthew resulted in missing the maximum flood extent. Similarly, Du et al. (2021) successfully fused surface fractional water cover (FW) and surface soil moisture (SSM) retrievals from Soil Moisture Active Passive (SMAP) L-band (1.4 GHz) brightness temperatures with rainfall forecasts for 1-day ahead forecasts of FW at 30m. Training a Classification and Regression Trees (CART) model implemented using Google Earth Engine (GEE; Gorelick et al., 2017), forecasts of FW inundation patterns were obtained by training the model on fractional water from Landsat data. While the study was innovative, it was limited to pixel-based prediction without considering contextual information.

Although contextual information could be included in the more traditional machine learning classifiers, e.g. through image segmentation and the inclusion of textural features (H. Song et al., 2019), algorithms such as CNNs are better suited to exploit the inherent spatial relationships (Rosentreter et al., 2020). In fact, Reichstein et al. (2019) suggest that the leveraging inherent contextual information through DL can support our understanding of Earth system science problems. The inherently contextual nature of CNNs also removes the need for feature engineering, which often needs further optimization, e.g. for choosing appropriate textures for classification as Dasgupta et al. (2018) demonstrate. In this context, the only study which attempted to simulate a SAR-scale observation for flood mapping using CNNs focused on generating the reference image for the commonly used change detection algorithms (Ulloa et al., 2022). Long-term SAR backscatter statistics were leveraged in this study within a DL-based spatiotemporal simulation framework using Convolutional Long Short-Term Memory (ConvLSTM) networks, to simulate a synthetic pre-image which aids flood detection the post-image is available. Practically, this approach still relies on the availability of the post-flood image for the actual flood detection and does not solve the problem of more frequent inundation monitoring.

Current capabilities in multi-source Earth Observation and deep learning, should theoretically allow mapping inundation at high resolutions even in the absence of a post-flood SAR observation, and this proof-of-concept study attempts to investigate the following challenging research questions:

RQ1.    Could multi-sensor multi-resolution remote sensing data enable flood inundation predictions at the S1 spatial resolution in the absence of a corresponding S1 image?

RQ2.    How well do the simulated flood probabilities correspond to the observed flooding during the event?

A purely remote sensing based approach is developed to answer these research questions, which leverages satellite-based coarse resolution frequent sub-/surface water observations and ancillary data as well as deep learning to provide high resolution flood inundation observations. Since, the S1 post flood image is not used as an input to generate the high-resolution flood inundations estimations, the proposed method demonstrates the feasibility of generating daily inundation products in the future for the first time. Specifically, a combination of pre-flood optical, SAR, and ancillary data (such as DEM and land cover), and a time series of post-flood hydrological variables (e.g. soil-moisture and precipitation) are utilized in this approach. A CNN model is trained using a flood map derived from a S1 SAR post-flood image, to predict inundation at the SAR spatial resolution. A concurrently captured cloud-free Sentinel-2 (S2) optical post-flood image served as an independent evaluation dataset for this study. Finally, a feature importance analysis is included to understand the model predictions better in terms of the input feature contributions.

# 2  Study Area and Data

## 2.1  Study Area

The region of interest (ROI) lies in southwest France and includes the city of Dax and its two main rivers, Adour and Luy, in the Département Landes and extends over ~1,577 km² (Figure 2a,b). Heavy rain from 10-13[th] December 2019 (see Figure 1a) caused a flash flood which inundated many areas in the ROI (Figure 2c-f, Copernicus EMS 2019). Streamflow data were obtained from the hydrology website of the Ministry of Ecology and Sustainable Development (Ministère de l'Ecologie et du Developpment Durable) at http://www.hydro.eaufrance.fr. A relatively down-stream gauge was selected to show the flood event as the upstream gauges mostly failed during the peak flow, however, the images were selected based on the evolution of the rainfall event upstream. The main reason for the choice of this ROI to the availability of concurrent S1 SAR and cloud-free S2 optical data capturing the flood inundation, acquired on the 15[th] of December. Due to the proof-of-concept nature of this study, the concurrent availability of these public datasets was vital to evaluate the posed research questions. The post-flood imagery available to this study is intended for use in the training and validation phases, but not as predictors which then limits the ability of the algorithm to the availability of these images (Figure 2d,f). It should be emphasized that this study region, was selected from over 150 flood events examined around the world, as the combination of concurrent cloud-free S2 optical and S1 SAR images, is rather rare.

## 2.2  Data Description

Multiple remote sensing datasets are used as predictors to estimate flood probabilities and described in Table 1 along with their main characteristics. The datasets are chosen such that the model as much information as possible regarding the pre-flood conditions and the incoming water during the flood event. Accordingly, the most recently acquired pre-flood S1 SAR and cloud-free S2 optical images, are provided as inputs to resolve the exact surface conditions just before the event. A DEM is additionally provided for elevation information as this controls the distribution of the incoming precipitation, and a Land Use Land Cover (LULC) map is included since this influences surface roughness that in turn accelerates or decelerates the inundation. As the LULC map production dates could significantly differ from the date of the flood event, the pre-flood S1/S2 images are provided to ensure that the model has the most updated information on surface conditions. At the same time these images may contain information on the inundation probability, thus, images close to the time of flooding should provide more information to the model than those with a longer temporal baseline. Soil moisture and precipitation time series for the date of the prediction are the main inputs characterizing the incoming water in the domain. Finally, using all the above-specified inputs, inundation is predicted for 15[th] December 2019, and validated against the unseen portions of the Sentinel-1 SAR based flood map as well as the independent Sentinel-2 image.

As inundation is a highly non-linear and dynamic process, the post-flood-images used for validation need to be concurrently acquired, to evaluate the chosen research questions in this study. For the pre-flood images used as inputs, however, the Sentinel-2 and Sentinel-1 images do not need to be concurrent, as they are just used as up to date information sources for the land cover which can be assumed to be static across several weeks. This makes it possible to use images acquired days or weeks before the flood, if cloud-free data closer to the time of flooding is not available. For this study area, however, even the pre-flood Sentinel-1 SAR and Sentinel-2 optical images were concurrently captured just before the flood event occurred on 10[th] December 2019. S1, S2, LULC and precipitation data are acquired from Google Earth Engine (GEE; Gorelick et al. 2017), while the DEM and soil moisture data are provided by Copernicus. Note that the datasets have notably different resolutions (see Table 1), and thus, the amount of information, i.e., the pixel counts vary significantly, with the high-resolution data displaying strong variance in clear contrast to the coarser datasets.

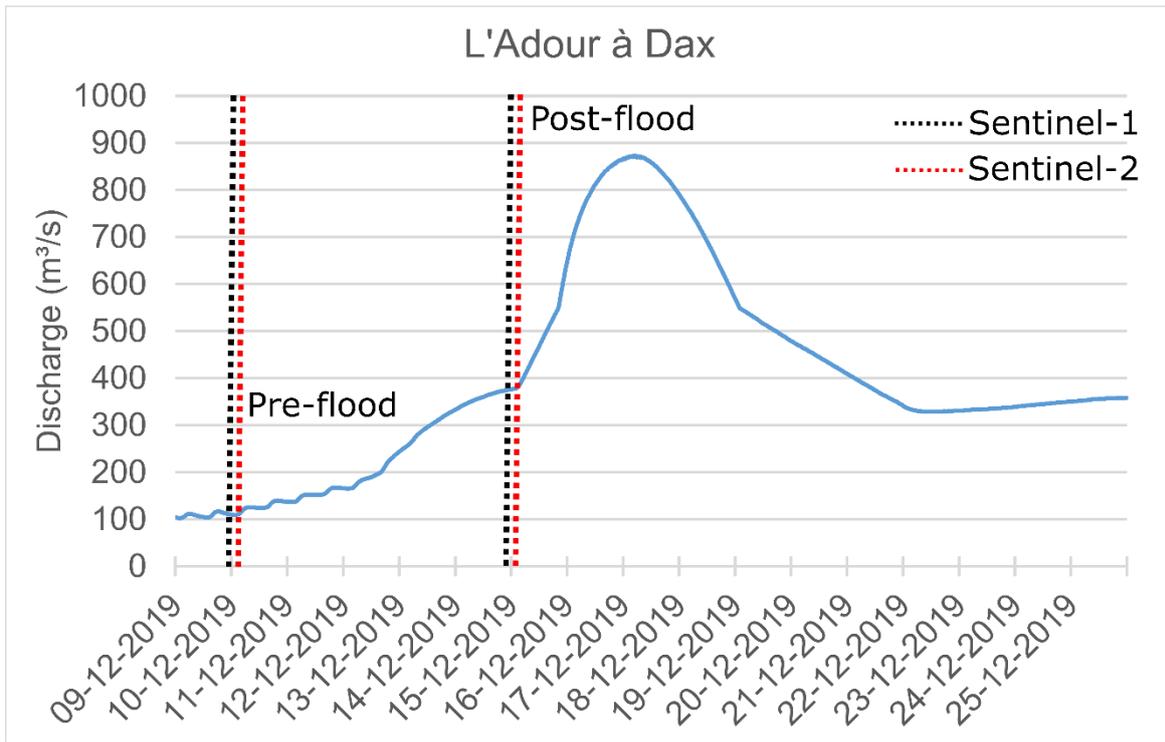

b)

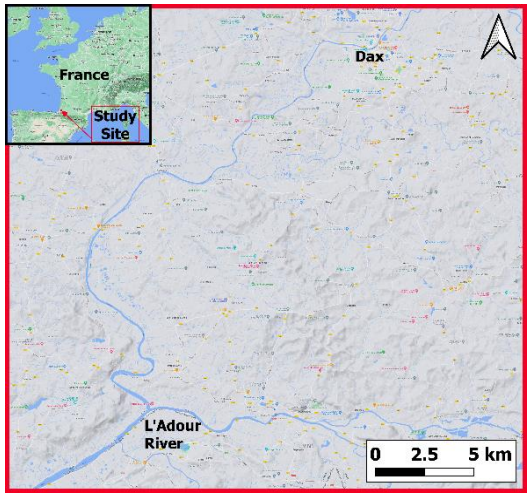

c)

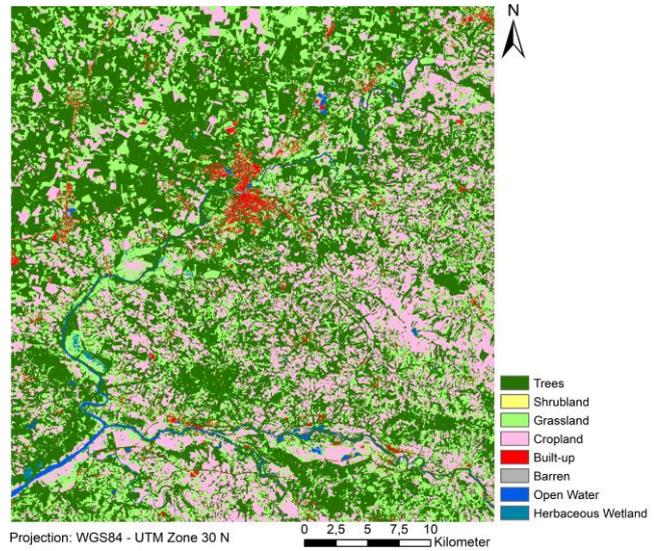

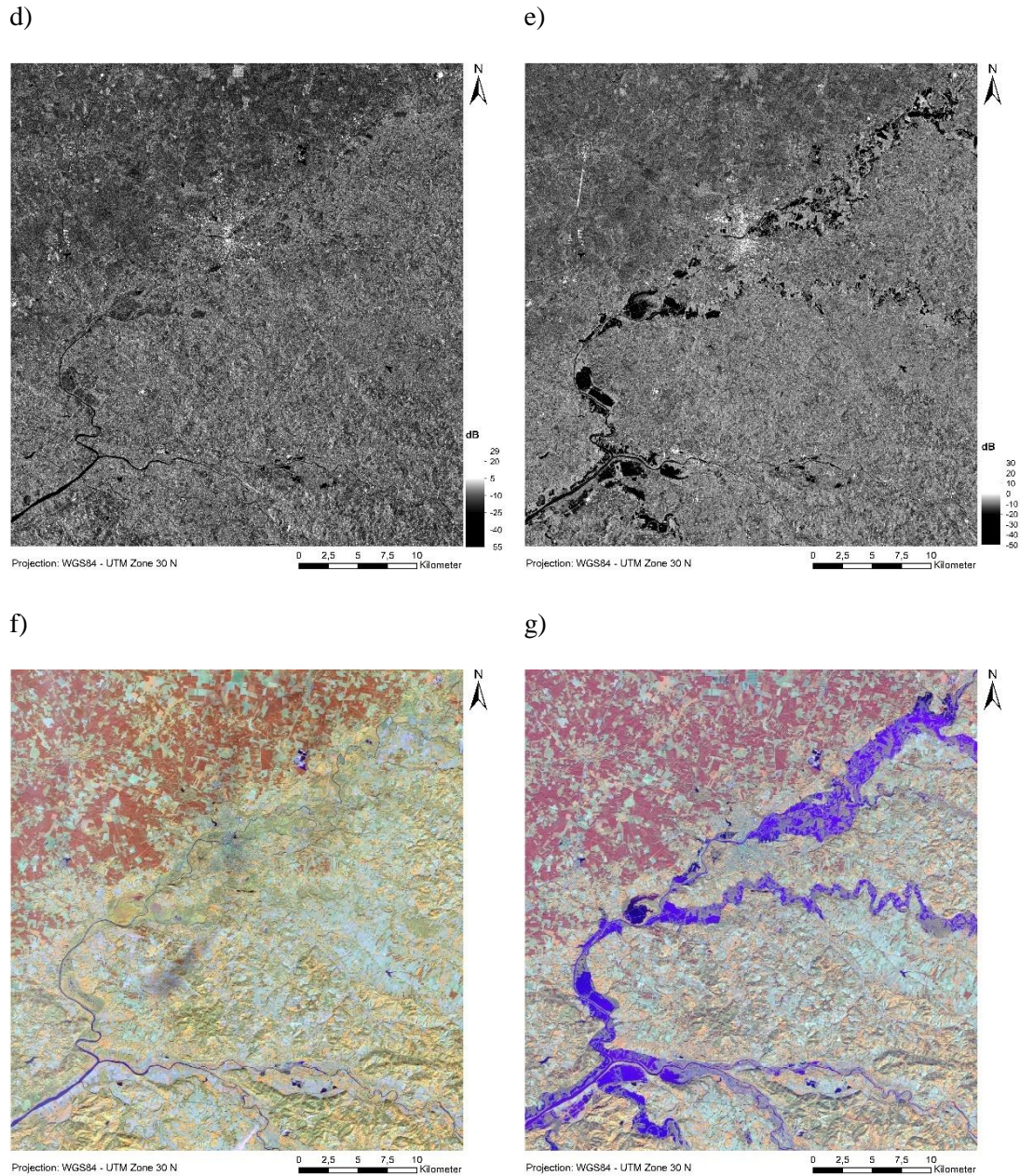

Figure 1: Study region and datasets used. (a) Hydrograph of the flood event, with discharge at the Arena Bridge in the city of Dax shown in cubic metres per second, along with temporal positions of the satellite acquisitions. Note: Satellite acquisition times are assigned as pre- and post-flood based on the upstream flood wave. (b) Study region, city Dax in south-west France. (c) ESA LULC (10 m) of the study area. Source: Zanaga et al. 2021. (d) Sentinel-1 pre-flood image in VV polarization, captured on 10/12/2019. (e) Sentinel-1 post-flood image in VV polarization captured on 15/12/2019. (f) Sentinel-2 pre-flood image as false-color-composite using (R: RedEdge4, G: SWIR1, B: Red) captured on 10/12/2019 (g) Sentinel-2 post-flood image as false-color-composite using (R: RedEdge4, G: SWIR1, B: Red) captured on 15/12/2019

Table 1: Summary of characteristics of the predictor dataset.

| Layer | Acquisition | Resolution | Units |
|---|---|---|---|
| **Sentinel-1** **VH, VV** | 10/12/2019 | 10m | dB |
| **Sentinel-2** **B, G, R, NIR** | 10/12/2019 | 10m | TOA reflectance |

| Sentinel-2 Red Edge (Bands 5,6,7,8A) + SWIR (Bands 11, 12) | 10/12/2019 | 20m | TOA reflectance |
|---|---|---|---|
| ESA-LULC | 2020 | 10m | categorical |
| COP30-DEM-Elevation | 2010 - 2015 | 30m | m |
| COP30-DEM-Slope | 2010 - 2015 | 30m | ° |
| COP-Soil-Moisture | 15/12/2019 | 0,25° | % |
| COP-Soil-Moisture-Uncertainty | 15/12/2019 | 0,25° | % |
| GPM-HQ-Precipitation | 15/12/2019 | 0,1° | mm/hr |
| GPM-IR-Precipitation | 15/12/2019 | 0,1° | mm/hr |
| GPM-CAL-Precipitation | 15/12/2019 | 0,1° | mm/hr |
| GPM-CAL-randomError | 15/12/2019 | 0,1° | mm/hr |

The S1 dual-polarized C-band (5.404GHz) SAR GRD data is provided by GEE as pre-processed tiles in decibel (dB) with a 10 m spatial resolution, including thermal noise removal, radiometric calibration and terrain correction using a DEM (Google 2021). Both polarizations, vertical transmit/vertical receive (VV) and vertical transmit/horizontal receive (VH) are used herein as predictors dataset, and the image was acquired during the ascending overpass. S2 data is acquired with 13 spectral bands using visible (VIS), near infra-red (NIR) and short wave infra-red (SWIR) regions of the electromagnetic spectrum. The Level 1C product is used in this study, which includes pre-processing steps such as radiometric correction, registration and orthorectification. As S2 spectral bands have different spatial resolutions of 10 m, 20 m and 60 m (Szantoi & Strobl 2019), coarser bands (60 m), containing information about aerosols, water vapor and cirrus were excluded from the predictors.

Recently, ESA published a global LULC product at a 10 m spatial resolution for the year 2020 (WorldCover) based on S1 and S2 data (Zanaga et al. 2021), which is also available in GEE (Zanaga et al. 2020). As the flood happened in December 2019, it is assumed that the LULC did not drastically change until 2020. WorldCover segments the earth's surface in eleven land-cover classes, including trees, shrubland, grassland, cropland, built-up areas, barren, snow or ice, open water, herbaceous wetland, mangroves, and moss and lichen, of which eight classes are represented in the ROI (Figure 1b). The Copernicus DEM (CopDEM30) at 30 m spatial resolution is used in this study as the elevation data source, which is based on SAR interferometry using images acquired by TanDEM-X from 2010 to 2015 (Airbus 2020). The elevation data is used to calculate additional topographic indicators such as slope, which are then added to the predictor dataset. While the data described above is expected to provide information on current surface conditions in the study area before the flood event, the soil moisture and precipitation datasets represent the post-flood conditions.

The antecedent surface soil moisture (SSM) provides information on soil saturation, with higher levels of surface wetness associated with rapid inundation following a rainfall event (Du et al. 2020). Many different SSM datasets exist but most high-resolution datasets depend on S1 or S2 data, again limiting their temporal resolution. The Copernicus ESA Climate Change Initiative (Copernicus Climate Change Service 2019), fused SSM product uses data from the active-microwave sensors ASCAT-A and -B and the passive-microwave sensors SMOS and AMSR2 with 1-2 day revisit times, enabling near-daily near-global coverage. The high revisit comes at the cost of spatial resolution, resulting in coarse grids of ~0.25°×0.25° (Kidd 2020), which leads to limited variability within the ROI. Post-flood Copernicus SSM (%) data, captured on the 15th December 2019 (date of prediction), is used herein along with the provided associated uncertainty layer.

Finally, precipitation data from the Global Precipitation Measurement (GPM) (Huffman et al., 2019), for the date of prediction is added as a predictor, assuming that the soil moisture already contains information on the prior rainfall. GPM provides 3-hourly precipitation observations on a global scale with a spatial resolution of 0.1° (~11×8 km), through data fusion of multiple passive-microwave satellite sensors (Huffman et al., 2015). While the pixel size is rather coarse with respect to the desired spatial resolution of the predicted flood maps, the variance of the GPM data is still high as compared to the soil moisture data. Only four of the ten available GPM bands are retained as inputs, including the merged microwave-only precipitation estimate (HQ), the Infrared-only precipitation estimate (IR) and the multi-satellite precipitation estimation with gauge calibration (CAL), along with its random error. The 3-hourly GPM products are aggregated to the daily scale to match the temporal frequency of the other inputs used in this study.

# 3   Methods

Since, most of the datasets, except for soil moisture and DEM (acquired directly from Copernicus), are taken from GEE, the datasets are already pre-processed (see Section 2.2 for details on the processing levels). The different coordinate systems and spatial resolutions, are first transformed to the WGS84 UTM (Zone 30N – based on the study site) coordinate system, resampled to a spatial resolution of 10 m, and stacked to form one multi-band image. Due to the contextual learning capabilities of CNNs (Zhang et al., 2016), no further pre-processing or feature engineering, is necessary. An overview of the proposed flood inundation estimation approach is illustrated in Figure 2. The reasons for choosing each processing step and its subsequent implementation are described in the following sections.

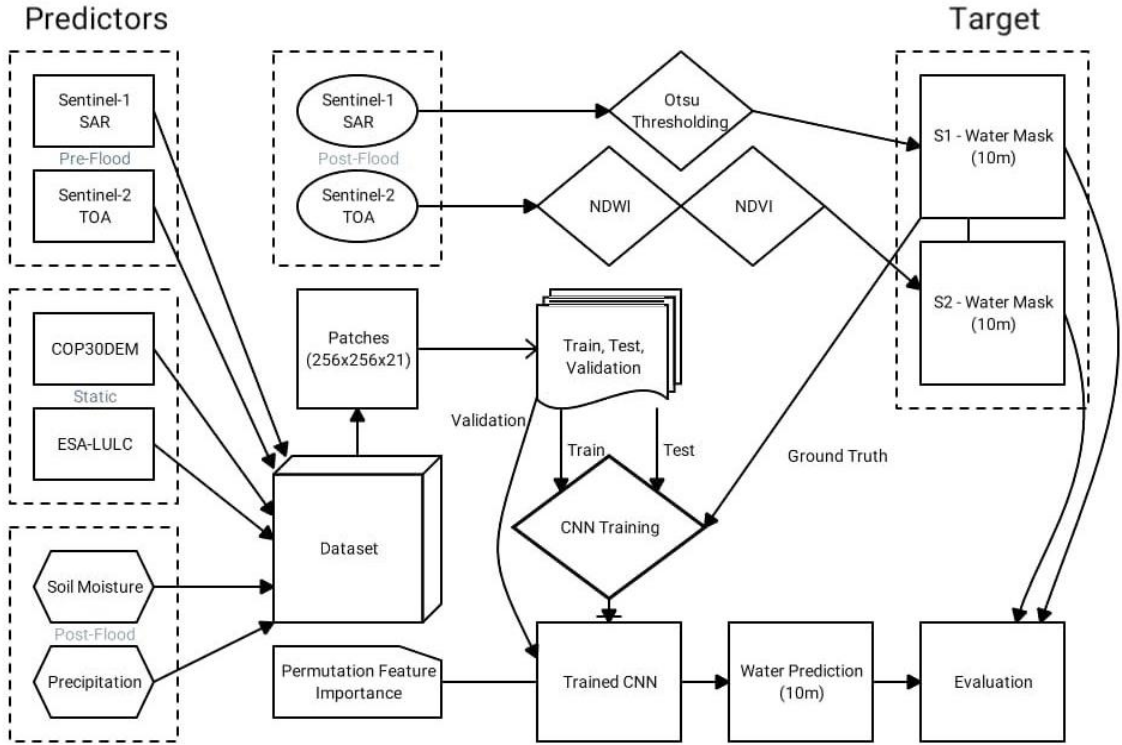

Figure 2: Flowchart describing the algorithm, predictors, target and ground truth acquisition.

## 3.1 CNN Model Design and Implementation

### 3.1.1 Model Architectures

Here CNN segmentation models are used, which were originally developed for segmentation tasks and thus can directly interpret EO images. The most popular CNN segmentation model i.e. the UNet, originally designed by Ronneberger et al. (2015) for biomedical segmentation, is chosen along with the SegNet model, which was developed with an aim towards scene understanding (Badrinarayanan et al., 2017). The UNet model uses skip connections between different blocks of layers to preserve the acquired feature maps (Figure 3). In contrast, the SegNet works strictly sequentially but reuses pooling indices from encoder for non-linear upsampling through the decoder thus reducing computational efforts. Here both architectures are slightly modified for the flood inundation predictions. The chosen UNet architecture was developed by deepsense.ai and reached the 4th place among 419 teams in the Kaggle competition "Dstl Satellite Imagery Feature Detection" (Nowaczyński et al. 2017). The task in the competition was to segment EO data into features, such as waterways, standing water, buildings, trees, tracks, roads, crops or vehicles (10 classes in total; Kaggle 2017). A SegNet model was also modified during a recent edition of the same competition for EO scene interpretation (Kaggle 2021).

As labelled patches are required for training such models, the original image with 21 input bands (Table 1) is sub-sampled to multiple smaller image patches. Fixed patch size of 256×256×21 pixels was used following Helleis et al. (2022b), since larger patch sizes are associated with better spatial feature encoding and densely connected high-level feature maps (Zhang et al., 2019). Recall that all predictor inputs were up-sampled to a pixel size of 10 m×10 m, resulting in multiple, overlapping patches, each containing 65,536 pixels (~6.55 km²; Figure 4a)

While the model's architectures are different they also share some properties. Both models are fully convolutional, not containing a fully-connected layer but generating the output with a convolution and a non-linear sigmoid function, which results in a probabilistic segmentation (Kabir et al., 2021). As patches are acquired with a size of 256×256 pixels and 21 bands, both models expect this input size to return similarly sized results, but with just 1 output band. The output pixel values vary from 0 to 1, where 1 represents inundation with high classification certainty, while 0 represents a dry pixel classified with high confidence. Both models use 3×3 filters with a 1×1 overlapping stride in the convolution layers and a 2×2 pooling window with a stride of 2×2 non-overlapping stride for down sampling the feature maps within the max pooling layers. Both models use the ReLU activation function in each hidden convolutional layer and the parameterization of the individual layers is also similar.

During the model development phase, multiple competing model structures were assessed, which revealed that batch normalization layers were critical. Since the input data are not explicitly normalized during the training and the units/magnitude differ notably, batch normalization layers must be included to normalize the data intrinsically. The original implementation of the SegNet architecture, includes paired convolution and batch normalization layers (Badrinarayanan et al. 2017), while the UNet architecture was modified similarly to include batch normalization layers absent in the original implementation (Figure 3, Ronneberger et al. 2013). The two models differ in the number and order of the layers (see Figure 3), as well as the skip connections used explicitly in the UNet. As each patch gets filtered and down-sampled multiple times, some high detailed feature information could be lost during the down- and up-sampling processes, e.g. when using the SegNet architecture. Conversely, the UNet concatenates the feature representations of the up-sampled and skipped portions in the decoder part of the model, through the in-built skip connections preventing such loss of information. The skip connections also enable the UNet to down-sample much deeper, generalizing the 256×256 pixels patch, with five blocks of convolutional, activation, max pooling and batch normalization layers, down to 8×8 pixels feature maps, before up-sampling them back to the original size (Figure 3a). The SegNet however only uses two blocks to down-sample the input, resulting in 64×64 pixels feature maps, which combined with the lack of skip connections could lead to poorer generalizability for both high and low level features (Figure 3b).

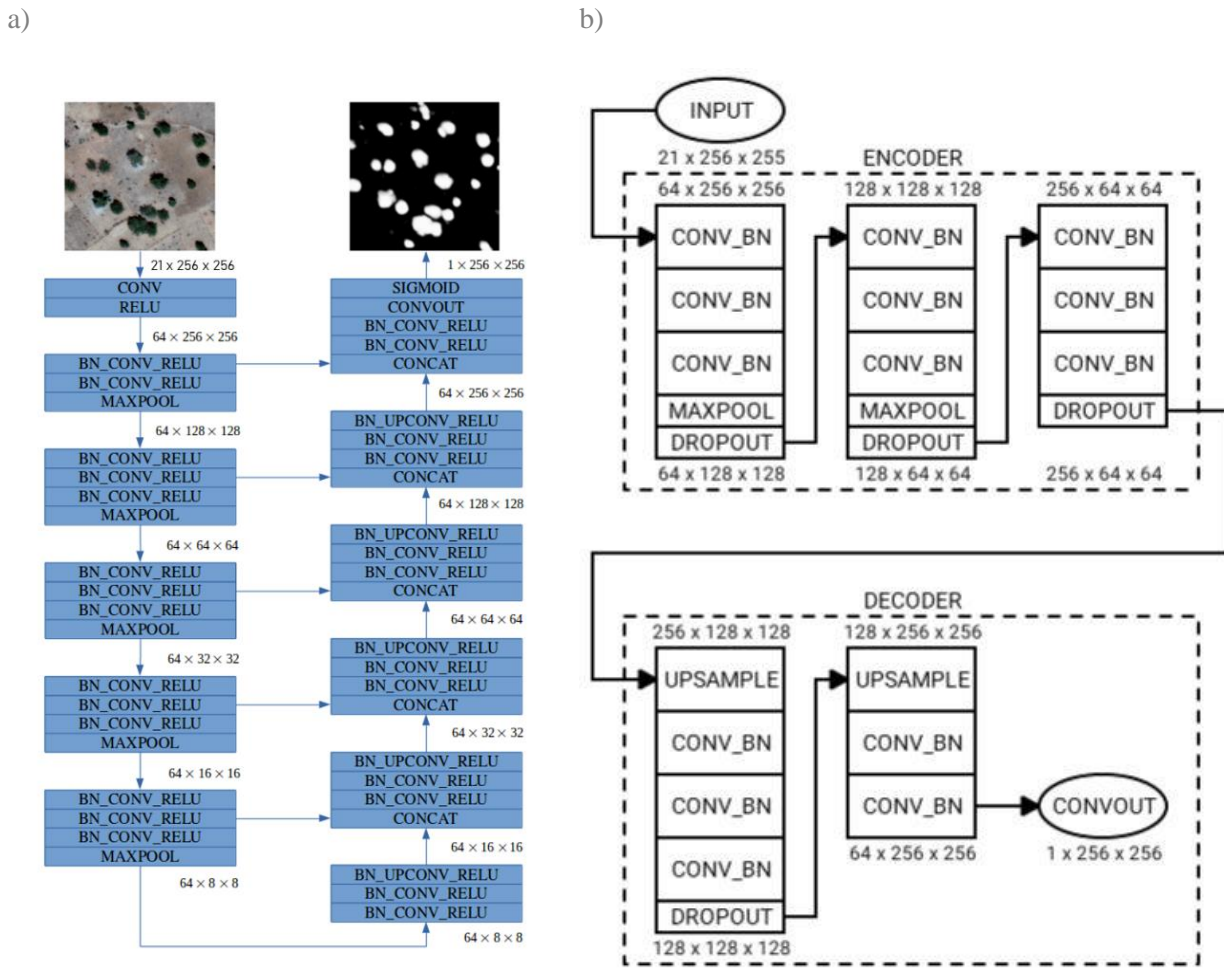

Figure 3: (a) UNet model from deepsense.ai constructed for satellite LULC segmentation. (b) SegNet model constructed for satellite LULC segmentation. Abbreviations: CONV = Convolution + Relu, BN = Batch Normalization. Modified after (a): https://deepsense.ai/deep-learning-for-satellite-imagery-via-image-segmentation/

### 3.1.2 Training Strategy

Both models are trained using a weighted binary cross-entropy loss, which is a widely used loss-function for flood mapping and imbalanced binary data (Bentivoglio et al., 2021; Muñoz et al., 2021). Since the class imbalance ratio varies within the training, testing and validation datasets, the weights to balance the loss are computed from the whole SAR-based water mask. Using the entire water mask yields the weight of ~0.5 for the background class (dry land) and ~15 for the flood water class. The models were trained using the Adam optimizer, which converges faster as compared to other optimizers due to its adaptive learning rate for each parameter (Taqi et al. 2018).

Besides non-trainable parameters both CNN models have a large number of trainable parameters. Larger number of trainable model parameters can lead to better performance in ideal conditions. However, such models also likely require a larger amount of training data to gain similar generalization capabilities, and can also lead to overfitting when using a small training dataset. The SegNet was thus modified to include some dropout layers, which are optional layers within CNNs used to prevent overfitting by dropping some of the connections in the network during training and only changing certain weights in each epoch (Badrinarayanan et al. 2017). The UNet does not include dropout layers in principle, and these were not added in during the modifications, as the skip connections are expected to optimize the model generalization capabilities (Figure 3a).

A detailed description of the generation of the training data is given in 3.2. The models are trained for up to 1000 epochs, in each of which the CNN is exposed to the full set of training data in batches, to reduce the computational requirements. Each batch consists of 10 randomly selected samples from the training dataset, which are used by the model to update the weights. At the end of each epoch, the test dataset is used to evaluate the model, enabling an assessment of the CNNs generalization capabilities on unseen data. An early stopping criterion is additionally used to prevent overfitting, i.e. if the loss function stops improving (patience parameter) on the test dataset for at least 50 epochs, the training stops and the model with the best testing score is restored (Muñoz et al., 2021). The CNNs were trained in a Python3 environment using Tensorflow and Keras, using two NVIDIA GeForce GPUs, each with approximately 24GB memory.

## 3.2 Training Data Generation

For supervised training of a CNN (see 3.1.2), the algorithm needs end-to-end labelled training data, to learn spatial patterns from its target. In this study, a binary water mask derived from the S1 SAR post-flood image is used as the target to train the CNN model, in order to be able to predict the inundation. However, it is worth underlining that the final model predicts flood inundation at the S1 spatial resolution in the absence of a corresponding S1 image for the target date, thus, making the predictions independent of the satellite revisit frequency.

### 3.2.1 S1-based Binary Water Mask Generation

Different approaches exist to detect water within SAR images, including histogram thresholding (Chini et al., 2017; Liang & Liu, 2020), coherence calculation (Chini et al., 2019; Pierdicca et al., 2014), classification (Helleis et al., 2022), region growing (Mason et al., 2014), fuzzy approaches (Dasgupta et al., 2018; Martinis et al., 2015) or change detection (Giustarini et al., 2013; Zhao et al., 2021). As the focus of the study is to evaluate the capability of the CNN model to predict high resolution inundation, simple thresholding based methods are used to generate the binary water masks (BWM) for the training (Mayer et al., 2021). While all the methods have been successfully used for SAR-based flood mapping, histogram thresholding is most commonly used (so far), as it is relatively fast, easy to implement, and automatable (Helleis et al., 2022; Liang & Liu, 2020). Thresholding methods depend on the assumption that flooded SAR image have bimodal histograms, typically containing two distinct distributions of water and non-water pixels, with the lower value maxima corresponding to water due to specular reflection (Liang & Liu 2020). Naturally, if the histogram does not display a significant contrast between the two classes, thresholding methods cannot be used. The bimodality assumption often fails if only a small portion of the image is flooded (Shen et al., 2019), in arid areas where dry sand displays similarly low backscatter (Martinis et al., 2018), or in windy areas where the water surface is roughened resulting in higher backscatter (Dasgupta et al., 2018). Since, the S1 post-flood image had a clearly bimodal histogram (Figure 4a), histogram thresholding was a viable option in this case to detect inundated areas. The Otsu method was chosen due to its simplicity and popularity, which automatically selects a global threshold to maximize the inter-class variance (Otsu, 1979). Only the VV band is retained for the inundation detection, as it leads to a more plausible representation of flooding (Clement et al., 2018), and an optimal threshold of -16.80 dB (Figure 4a), was selected based on the Otsu method.

a)

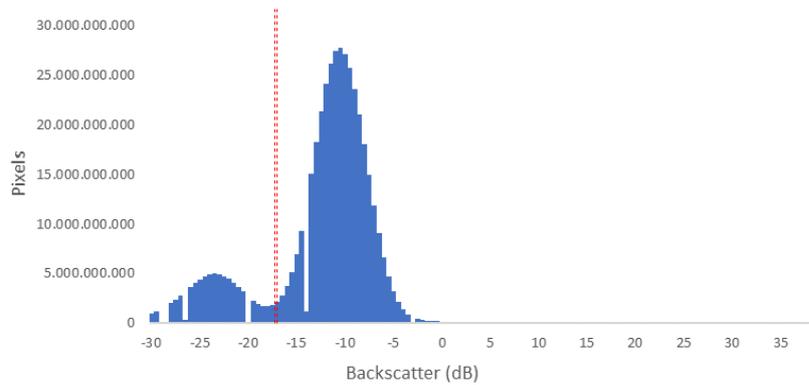

b)

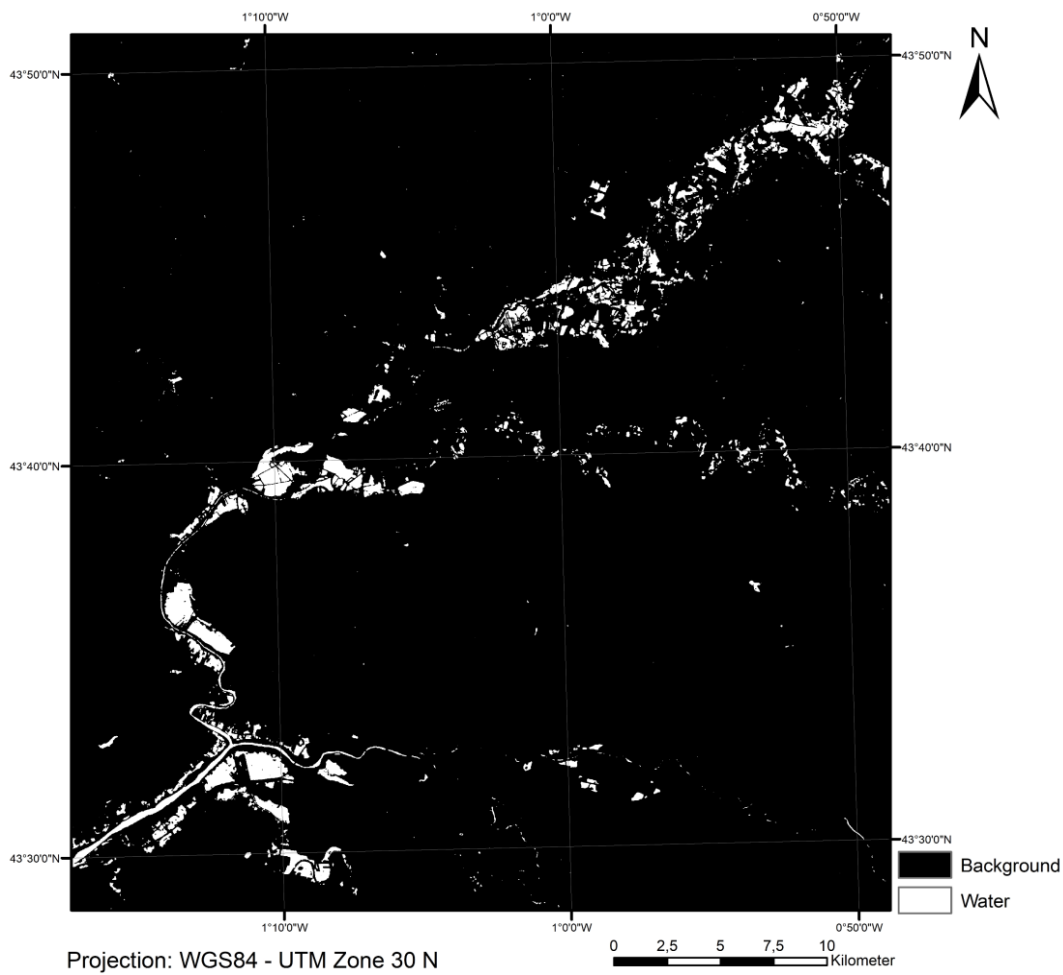

Figure 4: Training data generation using Otsu's thresholding method on the Sentinel-1 post-flood image VV band, where (a) shows the Sentinel-1 post-flood image histogram and (b) shows the Sentinel-1 derived binary water mask.

The side-looking geometry of SAR sensors often results in geometric errors such as radar shadows in steep areas, which are characterized by low backscatter and thus cannot be efficiently excluded from the water distribution by simple thresholding algorithms (L. Chen et al., 2020). Thus, as a necessary post-processing step (as the study area is mountainous), slopes greater than 7° were masked out from the analysis as water is expected to run off these regions (Wendleder et al., 2013). Further, to deal with the isolated dark pixels misclassified as water due to speckle noise in SAR images, a focal majority filter with a 3×3 moving window was implemented (Tetteh et al., 2021). Qualitative comparisons of the resulting binary water mask with the input SAR image, revealed most inundated (visually dark) areas to be correctly classified, although the majority filtering led to false negatives for some ephemeral streams. As smaller channels are typically harder to detect with the 10m resolution of S1 SAR images, due to lower widths which cannot be resolved by the sensor (Hardy et al., 2019). Still, the majority filtering step is retained as the mask retains most of the inundation, which was reasonably accurate to serve as labelled training data required for the CNN (Figure1d, Figure 4b).

### 3.2.2 Data Augmentation

As CNNs typically require massive amounts of training data in order to build generalization capabilities, data augmentation is rapidly becoming a standard processing step (Shorten & Khoshgoftaar, 2019), to expose the model (albeit synthetically) to diverse set of possible ground situations (Zhu et al., 2017). In fact, data augmentation has been shown to be of particular importance for SAR-based target recognition (Zhu et al., 2021), specifically helping to prevent overfitting, resulting in more robust and translationally invariant models (Ding et al., 2016). In the context of remote sensing based flood inundation estimation, data augmentation can also be used to reduce class imbalance within datasets (Waldner et al., 2019). As the S1-based BWM data contains only 3% flooded area is highly imbalanced (Figure 6c), augmenting the patches containing water, i.e. oversampling the under-represented class, particularly helps to ameliorate the class imbalance issue (Buda et al., 2018).

Due to the relatively recent uptake of deep learning in the field of flood mapping/modelling, there is no global consensus on optimal data augmentation approaches. While Bai et al. (2021) report slight increases in accuracy due to geometric data augmentation (flipping, rotating, and cropping), Helleis et al. (2022b) find radiometric augmentation (based on intensity or speckle noise structure) a better solution. Thus, simpler geometric augmentation schemes are chosen for this study (Yu et al., 2017), especially as the specific choice of the augmentation approach typically has a limited impact on testing scores (Helleis et al., 2022). The first data augmentation method is the stride, which defines the extent of overlap amongst neighbouring patches, and helps to improve the generalizability of the CNN learn by showing it the same area with different spatial contexts (Pereira & dos Santos, 2021). Using this method, 868 overlapping patches are acquired from the stacked input dataset (Figure 5a), along with their corresponding labels derived from the S1 BWM. Since, patches are acquired starting from the top-left corner of the stacked input image, areas at the bottom-right corner of the image are cut off if there are less than 256 pixels left (Figure 5b). Valid padding was chosen to rectify this problem, which essentially ignores the boundary data as there was no pressing need to retrieve a few pseudo-real patches at the borders (Wang et al., 2018).

Other augmentation methods are only applied to the training dataset, as flipping and rotation results in nonsensical imagery inconsistent with geometric properties of SAR sensors, as SAR image coordinates' represent the range and azimuth which are not arbitrary (Zhu et al., 2021). Thus, these processing steps are excluded from the test and validation datasets and just applied to the training dataset to mitigate the class imbalance (Zhang et al., 2021). Geometric augmentation for the training dataset was viewed as an acceptable compromise, as synthetic data have frequently been successfully used to train CNNs (Dahmen et al., 2019; Howe et al., 2020; Le et al., 2017). The training dataset was thus rotated and flipped repetitively by 90°, generating eight patches from each and allowing over-sampling of the flood class (Buda et al., 2018). Only patches which contain more water than the average patch (14 in total), were selected for the geometric augmentation, increasing the training set size from 86 to 198 patches (Figure 5a). Despite these measures, the training dataset remains highly imbalanced, as no patch consisting of >50% water exists (Figure 5c). Additional methods which counter class-imbalance, such as under-sampling the majority class, i.e. deleting patches which contain no water, are not used here, as the model should also learn to not over-predict in typically dry land. Another approach which could be used to increase the amount of water in each patch is the use of smaller patches, however, this would reduce the information on the spatial context provided by larger patches (Zhang et al., 2019). However, the class imbalance can also be countered within the CNN model structure, by using a weighted loss function during the training or controlling class-specific attention mechanisms (Johnson & Khoshgoftaar, 2019).

a)

Patch Properties

| Patch-Size | 256x256 px |
|---|---|
| Overlap | 128x128 px |
| Resolution | 1 Pixel = 10 m |
| Patch-Area | = 65.536 px |
| | = 6.553.600 m² |
| | = 6,55 km² |

Study Region

| Height, Width | 3968, 3584 px |
|---|---|
| Area | ~ 1422 km² |
| Rows, Columns | 31, 28 patches |
| Samples | 868 patches |

Number of Samples in each Dataset

| Dataset | Patches |
|---|---|
| Training | (86) → 198 |
| Testing | 89 |
| Validation | 40 |
| Excluded | 653 |
| Total | 868 |

b)

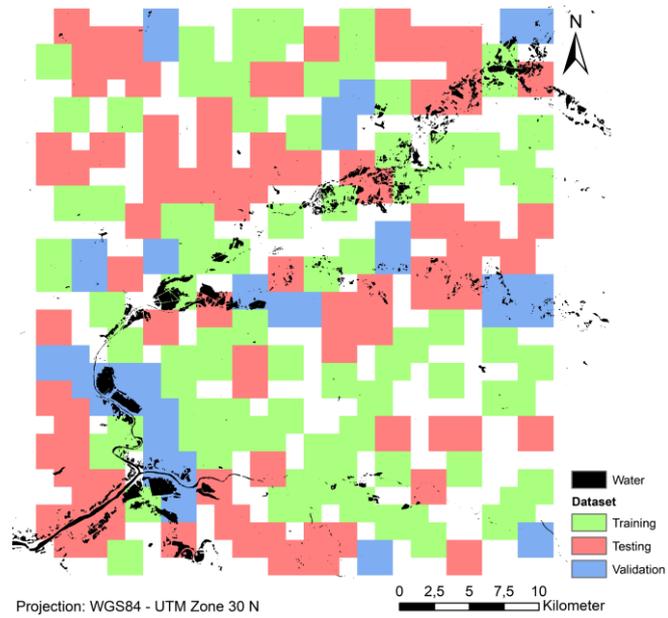

c)

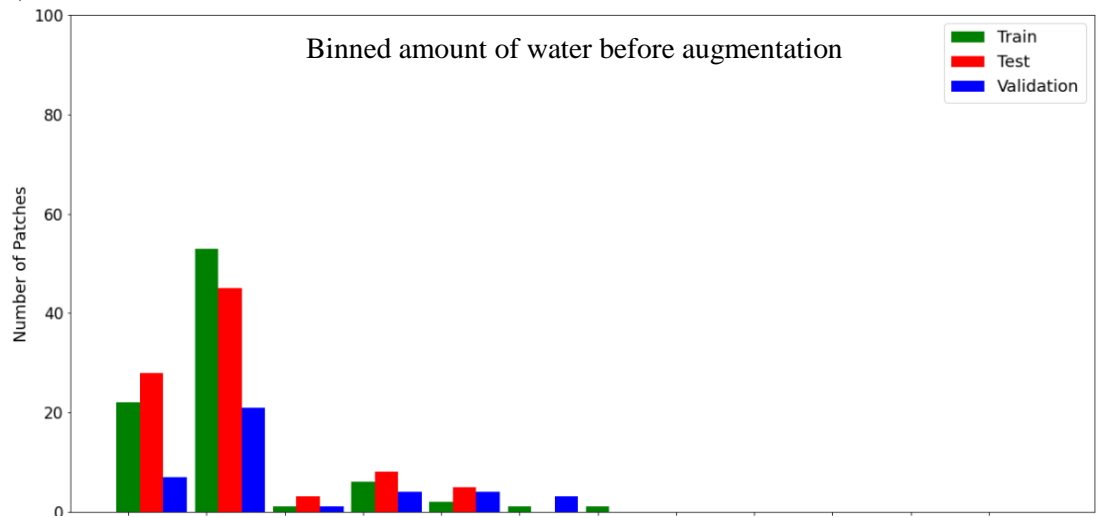

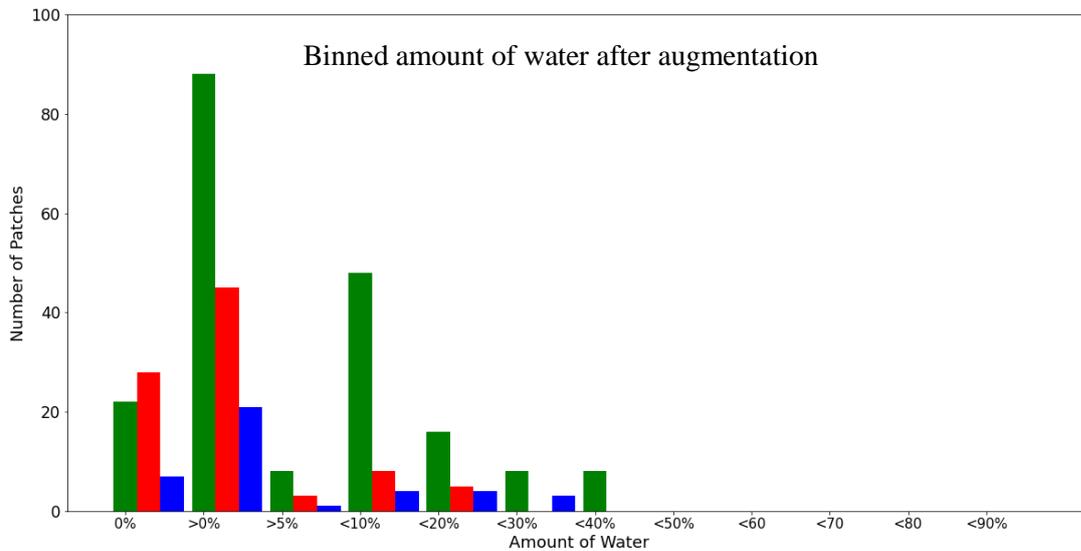

Figure 5: Datasets. (a) Study area properties including area and number of samples. (b) Spatial distribution of train, test, and validation datasets after random splitting and eliminating overlapping patches. (c) Binned, relative amount of water in drawn patches for each dataset before and after augmentation.

### 3.2.3 Training, Testing, and Validation Data Splits

The training data generated above for supervised learning of the CNN, is first split into training, testing and validation sets. Thus, the training split is used for the actual model training i.e., to adjust model weights/parameters, the testing dataset is used to assess model performance at the end of each epoch (Song et al., 2020). The test dataset prevents overfitting the model and increases overall generalizability, by monitoring the model's performance on "unseen" data excluded from model training (Bentivoglio et al., 2021). Finally, a third split of this dataset is kept completely hidden from the model during the training phase and serves as the independent validation. Thus, the overall augmented training dataset was randomly split into training (~40%), testing (~40%) and validation (~20%), by applying a seeded random permutation (to enable reproduction) and tracking the index (representing spatial location) of each individual patch. As overlapping patches were originally generated and since they cannot be shared across the train/test/validation datasets, patches which overlap into another data split are excluded, i.e. if a training set patch overlaps with one in the validation set, it is excluded from the validation set. Each dataset's number of patches, randomly selected spatial distribution, and content of water per patch in terms of %age of pixels in each (before and after the augmentation) are shown in Figure 5.

## 3.3 CNN Model Validation

The validation process is split into three parts. The first part is an internal model validation to and evaluate the training procedure and how well the competing models predict on the test dataset. The second part evaluates the trained CNN to predict the independent validation dataset, which is not used during training of the CNN models and is thus a reliable resource about its performance on unseen data. Finally, the trained models are evaluated against the Sentinel-2 based flood mask to understand the overall performance. The generation of the Sentinel-2 based flood mask is described on 3.3.1, while performance metrics for the validation are given in 3.3.2.

### 3.3.1 Sentinel-2 MSI Flood Mask Generation

A cloud-free Sentinel-2 (S2) post-flood image was used for validation, with index-based methods used to interpret the image into inundated area (Acharya et al. 2018). The S2 image not only provided a completely independent dataset for validation since it was not used elsewhere, but it also provided the advantage of being spatially and spectrally uncorrelated from the model which was completely SAR-based. Here, the normalized differenced water index (NDWI) and the normalized differenced vegetation index (NDVI) are used (Figure 6). While the NDVI is typically used to detect vegetation, with high NDVI values associated with green/healthy vegetation, negative NDVI values similarly can be interpreted as water (Cao et al. 2021). The NDVI is thus used to complement the information provided by the NDWI (positive values indicate water), such that only pixels with a negative NDVI and a positive NDWI are detected as water and used to generate a binary water mask (Acharya et al. 2018). The water mask is smoothed using a majority filter to exclude isolated, misclassified pixels. (Figure 6b, Tetteh et al., 2021).

Compared with the SAR-derived water mask, the optical water mask shows a lot more inundation, which is somewhat expected, due to sensor and resolution differences. While the Sentinel-1 water mask shows about 3% or 52 km² of the image to be inundated, the Sentinel-2 mask sees almost twice as much, indicating about 6% or 95km² of the area within the image to be water. Due to the absence of field data, the accuracy of the individual masks cannot be evaluated at this time. However, this could have some implications on the model performance assessments, as the "water-poor" SAR-based mask is used for training while the "water-rich" optical mask is used for validation.

a)

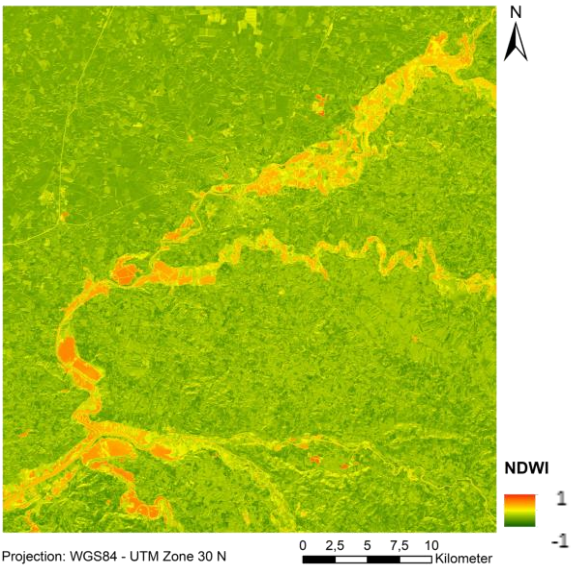

b)

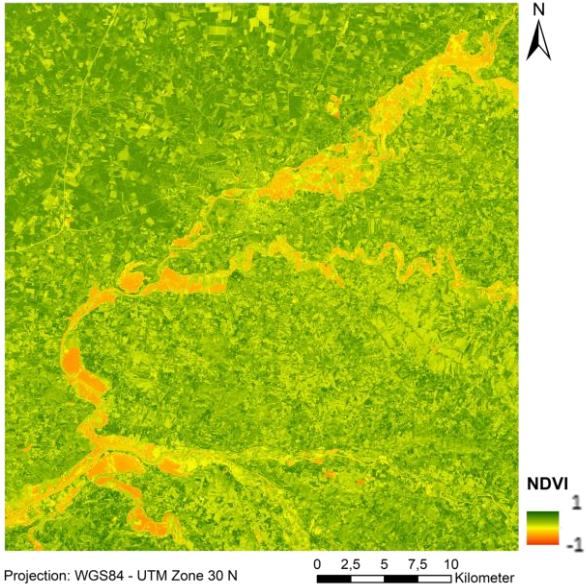

c)

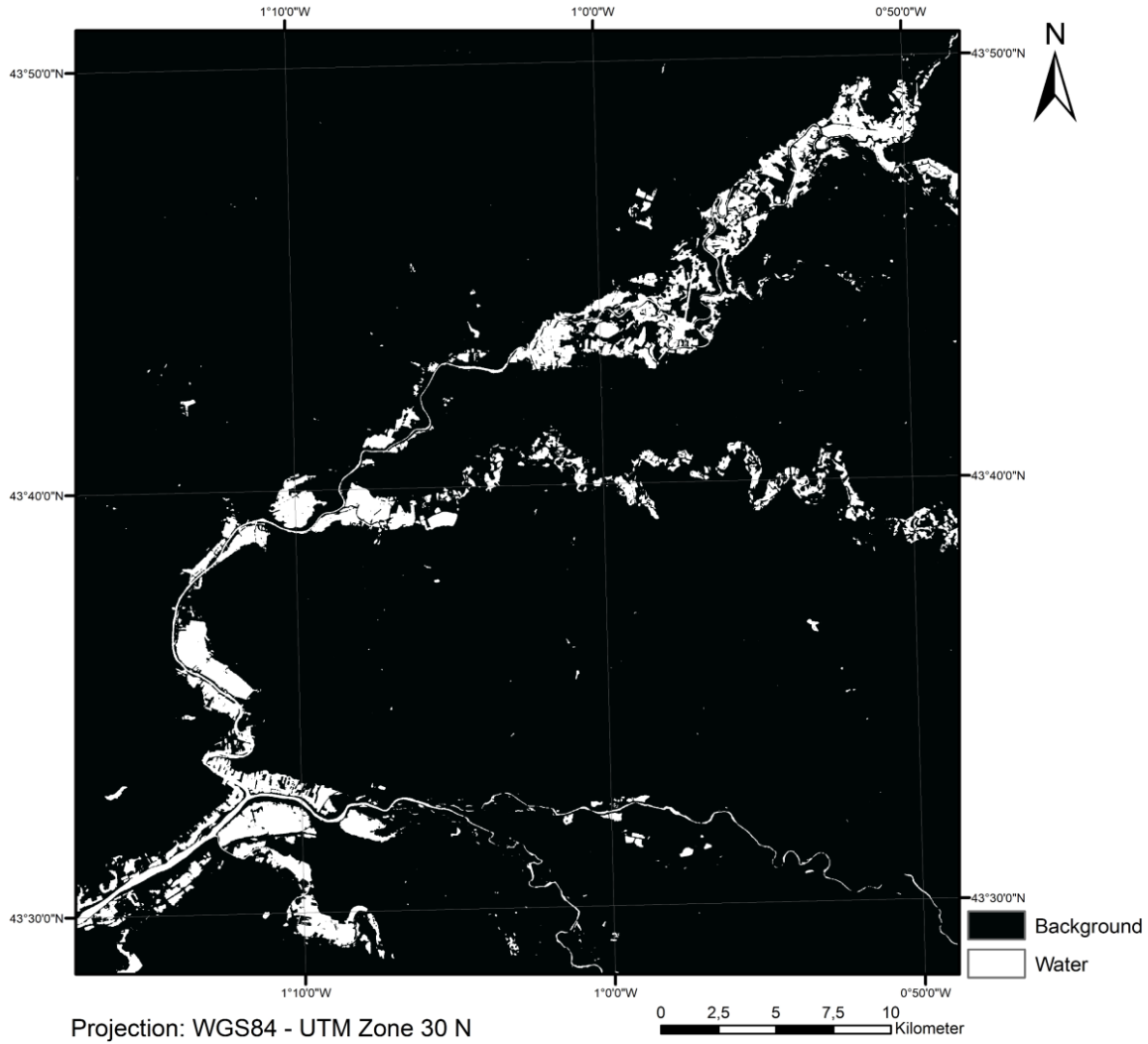

Projection: WGS84 - UTM Zone 30 N

Figure 6: Ground truth acquisition using a combination of NDWI and NDVI with the Sentinel-2 post-flood image. (a) NDWI = (Green-NIR)/(Green+NIR). (b) NDVI = (NIR − Red)/(NIR + Red). (c) Sentinel-2 derived ground truth.

### 3.3.2  Performance Metrics

Besides the weighted cross-entropy loss (see model training 3.1.2), different metrics including classification accuracy, precision, recall, F1-Score, area under the curve (AUC) of the Receiver Operating Characteristics (ROC) and AUC of the precision-recall curve (PRC) were also evaluated at the end of each epoch (S. Dong et al., 2021). These metrics were chosen to allow comparability of outcomes across the field, as they are the most popularly used in deep learning for flood mapping applications (see Figure A1 in Bentivoglio et al., 2022). Results were evaluated with the AUC of the PRC (PR-AUC), which is closely related to the F1-Score and does not account for True Negative samples i.e. correctly classified background pixels, thus, being better suited to highly imbalanced data (Z. Dong et al., 2021). Furthermore, the continuous probabilistic results needed to be binarized using a probability threshold, to calculate the confusion matrices for both CNNs' predictions against the Sentinel-1/2-based water masks. The confusion matrixes are used to calculate the User's and Producer's Accuracy (UA and PA) for each class, along with their confidence intervals computed as a function of their variances based on the area proportions observed in the sample (Olofsson et al., 2014).

## 3.4   Feature Importance Analysis

Lastly, the importance of different input features was also examined. Despite the recent popularity and success of CNNs, they essentially still remain "black box" models, with no direct way to implement and assess the importance of individual input features. Further, it is not possible to exclude specific layers from the input stack to assess their value when predicting using a trained CNN, as an input size identical to what was used in the training is required. Otherwise, the CNN needs to be retrained from scratch with the different input size, possibly resulting in a totally different weighted model, and thus, preventing an evaluation of feature importance in the original model (Alzubaidi et al., 2021). The most common approaches for feature importance analysis (or interpretable machine learning; IML) are the permutation feature importance (PFI; Breiman, 2001) and the conditional feature importance (CFI; Strobl et al. 2008). Both methods were originally developed for implementation within Random Forest classifiers, but model agnostic versions have recently been developed to extend the applicability of these methods to CNN models.

While PFI uses a perturbed version of the feature of interest sampled from the marginal distribution for replacement, CFI preserves the conditional distribution with respect to the remaining features during the perturbation. PFI, thus, provides a measure of the overall model reliance on a particular feature of interest, while CFI quantifies the unique contribution of the feature given all remaining features, i.e. excluding correlated information (König et al., 2021). The PFI method is practically implemented by reallocating pixels by changing their position but keeping their original values (Fisher et al., 2019). However, given the coarse resolution of the satellite hydrology datasets (soil moisture and precipitation) used herein as input features, where each patch usually only contains 1-2 different values, this random permutation approach might result in an input identical to the original making this method unsuitable for this study. Therefore, a different method is proposed to evaluate the CNN's feature importance.

First, the minima and maxima of the entire input image (for all 21 layers) were calculated. This original range of values was then utilized to generate random noise and subset the images into patches of the required size. These manipulated datasets were then used for prediction with the pre-trained CNNs and evaluated by comparing the PR-AUC of the CNN model fed with the original dataset to the CNN model fed with the manipulated datasets. Naturally, replacing an input feature important to the model predictions with noise layers, should result in a greater drop in the chosen accuracy metric as compared to a redundant feature. Using this principle, the relative importance of the input features can be ranked using the change in accuracy resulting from their replacement with noise.

# 4   Results and Discussion

## 4.1   Training Evaluation

Evaluating the training quality of the CNNs is an essential first step to understand the strengths and weaknesses of the models. To ensure unbiased comparability among the CNNs, the model training procedures were identical by design, including identical training data and hyperparameters such as the loss function and weights, optimizers, epochs, and the early stopping criteria (Section 3.1).

### 4.1.1   UNet

The UNet trained for 138 epochs, after which the early stopping criterion was triggered, as the testing loss did not improve for 50 consecutive epochs and the model from the epoch with the lowest testing loss was restored (88[th]). Each epoch took approximately 3s to run on the available computing resources, as each of 20 steps contained within required a compute time of 130ms. Figure 7 shows the evolution of the various loss and evaluation functions considered herein with each consecutive epoch, as well as the loss function values at early stopping. The training progresses rapidly within the first 10 epochs due to the use of the Adam optimizer, which has been shown to converge considerable faster than other optimization algorithms for deep CNNs (Kingma & Ba, 2015). Further training progresses slowly as expected, showing only minor improvements with increasing epochs. The peaks observed in the loss function curve (blue line), correspond to the optimizer overcoming a local minima during the gradient descent backpropagation (van der Meer et al., 2022).

Due to the huge class imbalance (only 3% foreground), the area under the AUC-ROC highly overestimates the models accuracy (Erickson & Kitamura, 2021). The (overall) accuracy metric is less sensitive to changes compared to loss and the precision and recall related F1-Score or the PR-AUC, as it includes the correctly predicted background values in the metric calculation which overpowers variations in false alarms, misses, and hits (Stephens et al., 2014). While the training loss continues to improve after the early stopping is applied, which is expected as the model gets more familiar with the cases included in the training set and starts overfitting. However, the testing loss ceases to improve as the model is unable to generalize on unseen data (

Figure 8). The decay in the testing loss is initially well-aligned with the training loss (blue line), often even lower than the training loss until the 70<sup>th</sup> epoch, but starts to exhibit strong oscillations afterwards. After the early stopping criterion is reached and the model starts to overfit the training data, the testing loss increases as expected (orange line).

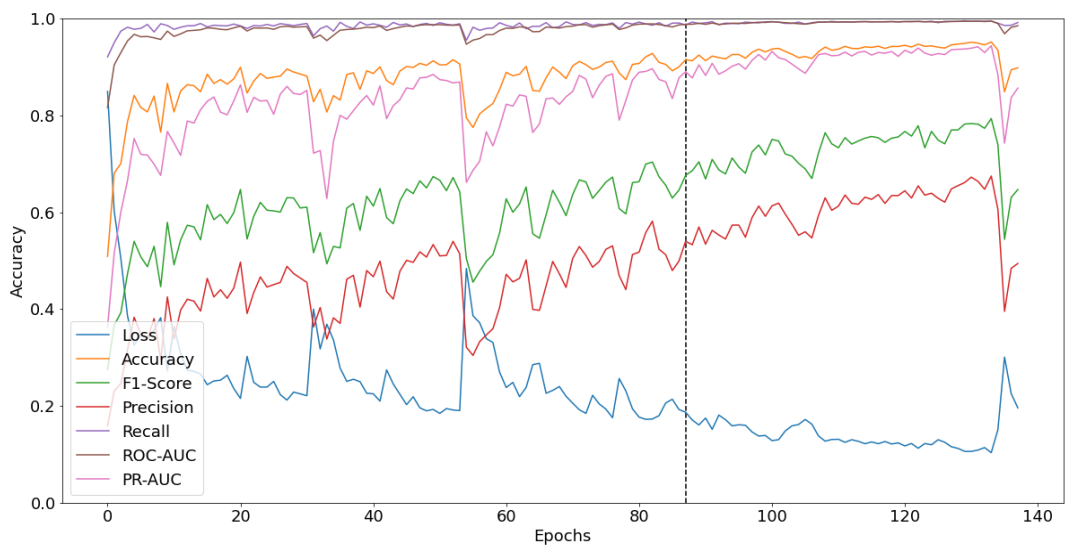

Figure 7: Plot of the UNet training accuracy variations during the training process for each loss/evaluation function used within the model (solid colored lines) with the early stopping epoch shown in the black dashed line.

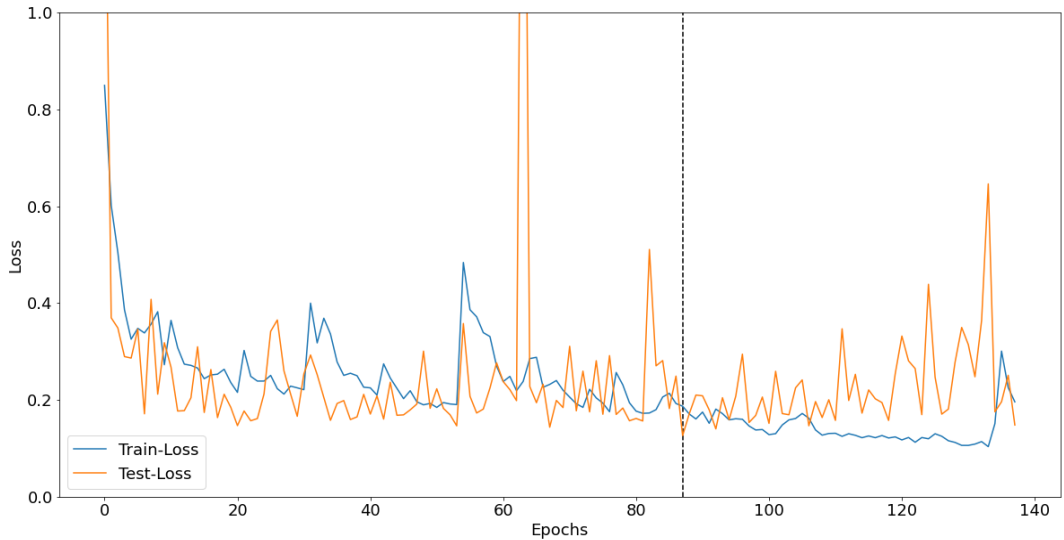

Figure 8: UNet weighted binary cross-entropy loss for train and test dataset with early stopping.

### 4.1.2 SegNet

The SegNet finished training after 129 epochs, slightly faster than the UNet (138 epochs), but the loss and accuracy metrics did not improve as fast as in the case of the UNet (Figure 7, Figure 9). Each step took a compute time of 150ms and each epoch ran in approximately 3s, indicating similar training time requirements to the UNet, despite the SegNet having almost twice as many trainable parameters than the UNet. While the training loss and other metrics continue to improve, the testing loss appears to be unable to converge, which could be caused by the lack of adequate training data for the massively increased number of parameters. Due to the use of early stopping a reasonably well performing model can be chosen, however, this seems to be a chance occurrence rather than a consequence of the knowledge gained through the training (Figure 10). It is also worth noting that the training loss values for the SegNet at the early stopping epoch are visibly higher than the UNet, while the testing losses are nearly similar, which can also be attributed to the inadequacy of the training data for the SegNet model.

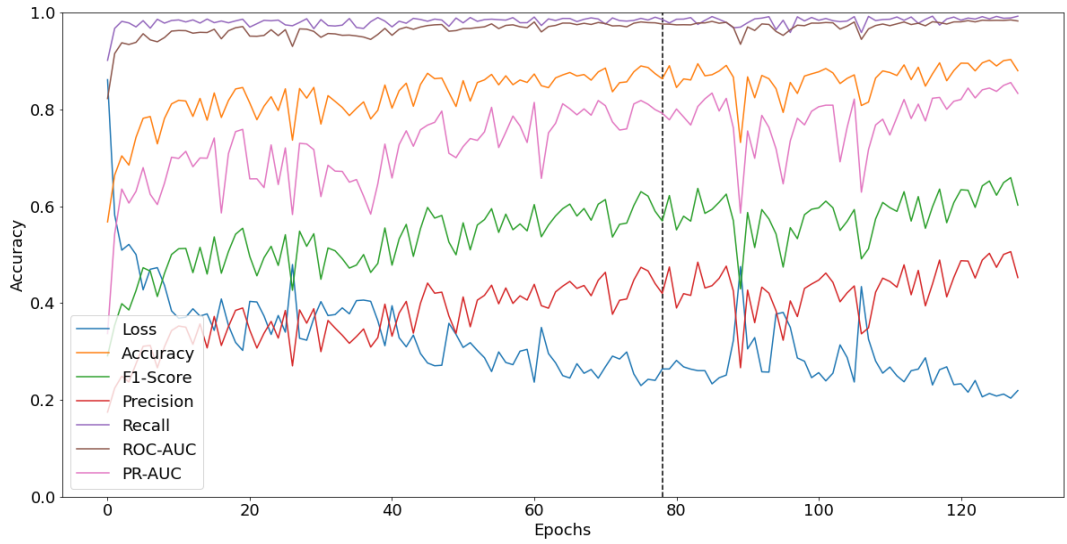

Figure 9: SegNet training accuracies for each epoch with early stopping.

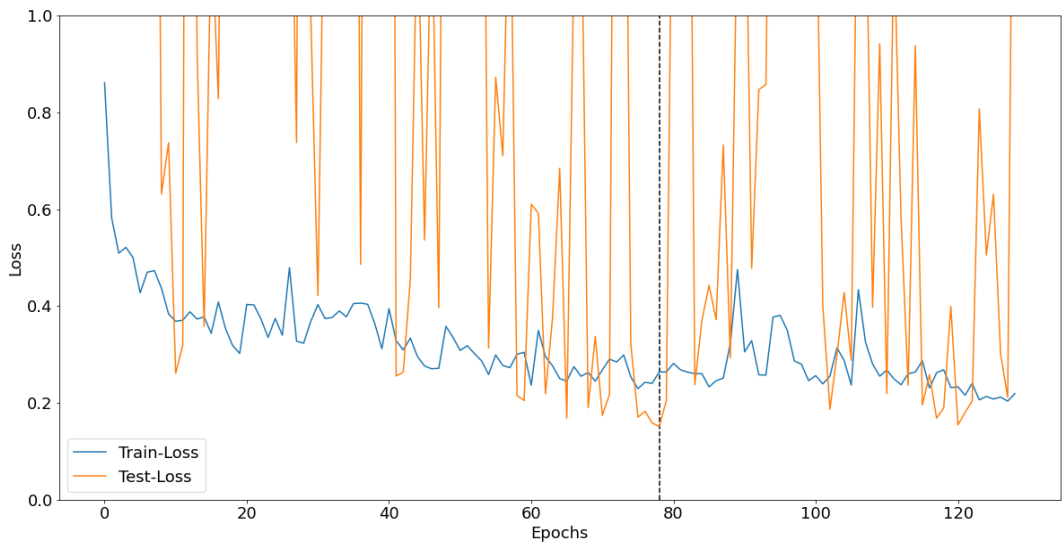

Figure 10: SegNet weighted binary cross-entropy loss for train and test dataset with early stopping.

## 4.2 Model Validation

a)                                            b)

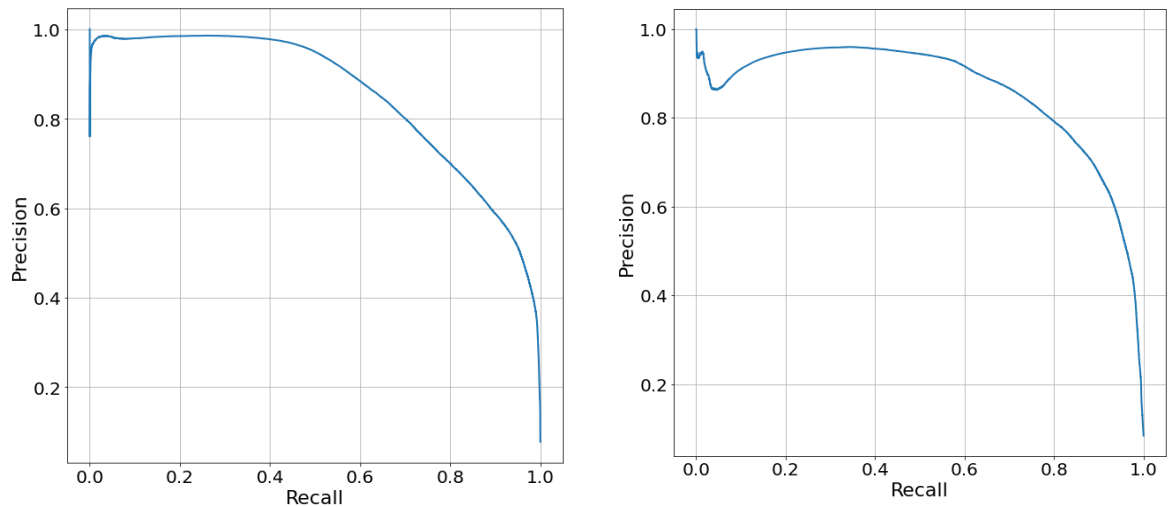

Figure 11: Plots showing the Precision-Recall Curves of the UNet (a) and the SegNet (b).

This section addresses the evaluation of the trained CNNs against the independent validation set (blue patches, Figure 6(b)), subset from the Sentinel-1 based water mask (Section 3.2.3) as well as the Sentinel-2 based water mask (Section 3.2).

### 4.2.1 Model Precision and Recall

Precision and recall are calculated to generate and evaluate the precision-recall curve. Recall represents the fraction of observed inundation which has been detected/predicted by the trained model, while precision measures the fraction of correctly classified inundation with respect to all areas classified as inundation. The PR-AUC values of both models are similar, with ~0.860 for the SegNet and ~0.854 for the UNet (Figure 11). Despite nearly identical PR-AUC values, closely examining the curves reveals visible differences. While the UNet has higher precision values at a recall of up to ~0.5 after which there is rapid decay, the SegNet exhibits higher precision values at recall values between ~0.5 to ~0.9. At low recall values, only those pixels are classified as inundation where the model is confidently able to identify them as flooded. Even within these pixels which are almost certainly inundated, the SegNet seems to make rather erroneous predictions as compared to the UNet. The low precision values observed for the SegNet even at low recall, indicates that the SegNet could wrongly be confident about the flooding status of a pixel. On the other hand, a recall value of 1 signifies that all inundated areas are detected. While both models were able to capture all inundates areas, this comes at the cost of the precision as more background pixels are classified as water leading to false alarms. The UNet however outperforms the SegNet at lower recalls with a noticeably higher precision, indicating that the inundated pixels identified by the UNet with high confidence are actually water.

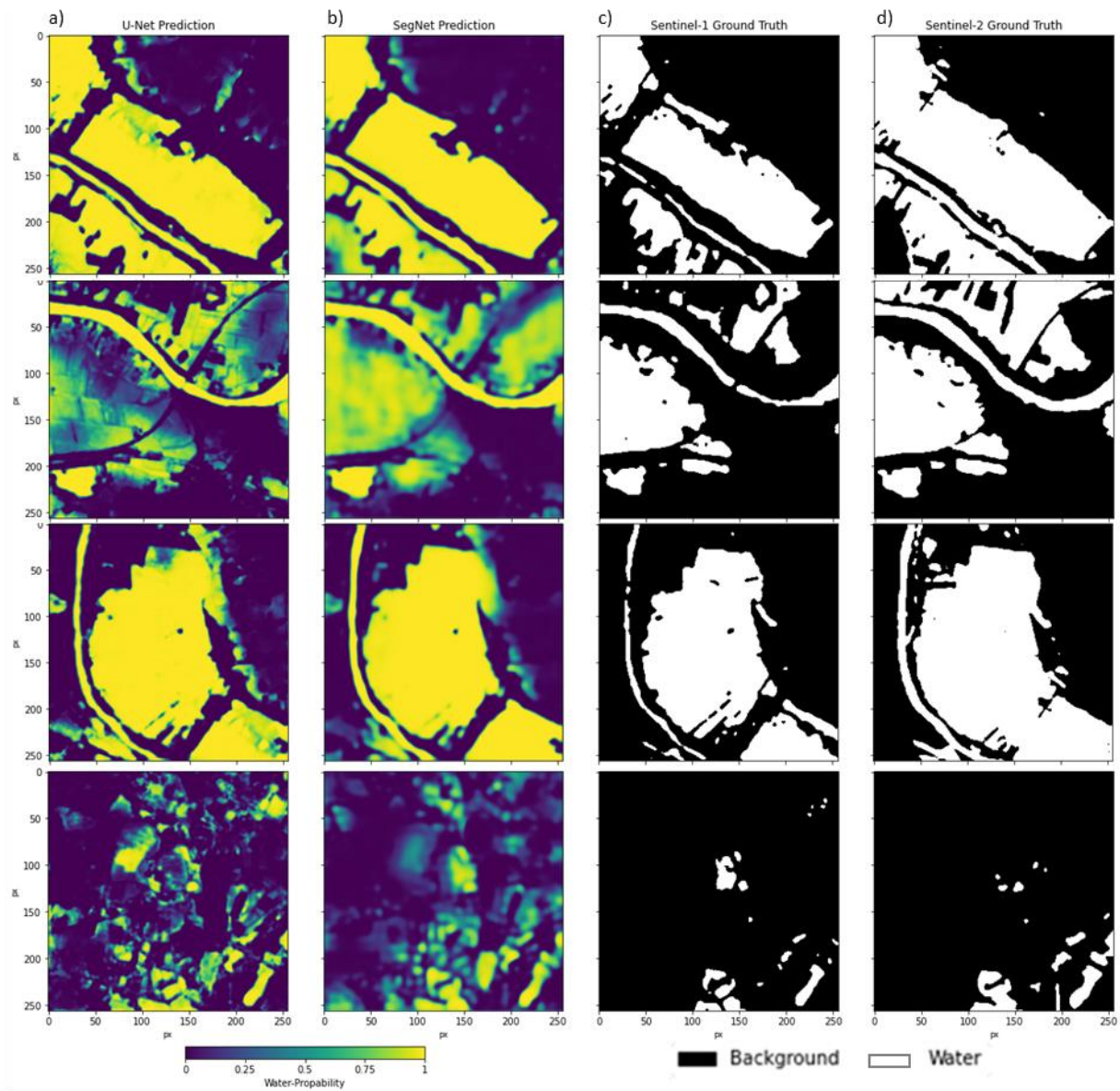

Figure 12: Predicted patches. (a) UNet prediction. (b) SegNet prediction. (c) Sentinel-1 ground truth. (d) Sentinel-2 ground truth.

Both classifiers are able to predict inundation at the Sentinel-1 resolution, showing similar results closely related to the Sentinel-1/2 Water Masks (Figure 12), despite the use of the coarse input datasets. The results show that both classifiers are able to capture the full extent of inundation, on a very detailed level, with only minor misclassifications using the proposed input data (Figure 12). Predictions generated by the SegNet's appear to have lost some finer resolution details, probably caused by the up-and down-sampling architecture as discussed in Section 3.1.1. The UNet, making use of the skip-connections, is able to preserve high-leveled details better than the SegNet as expected (Figure 12).

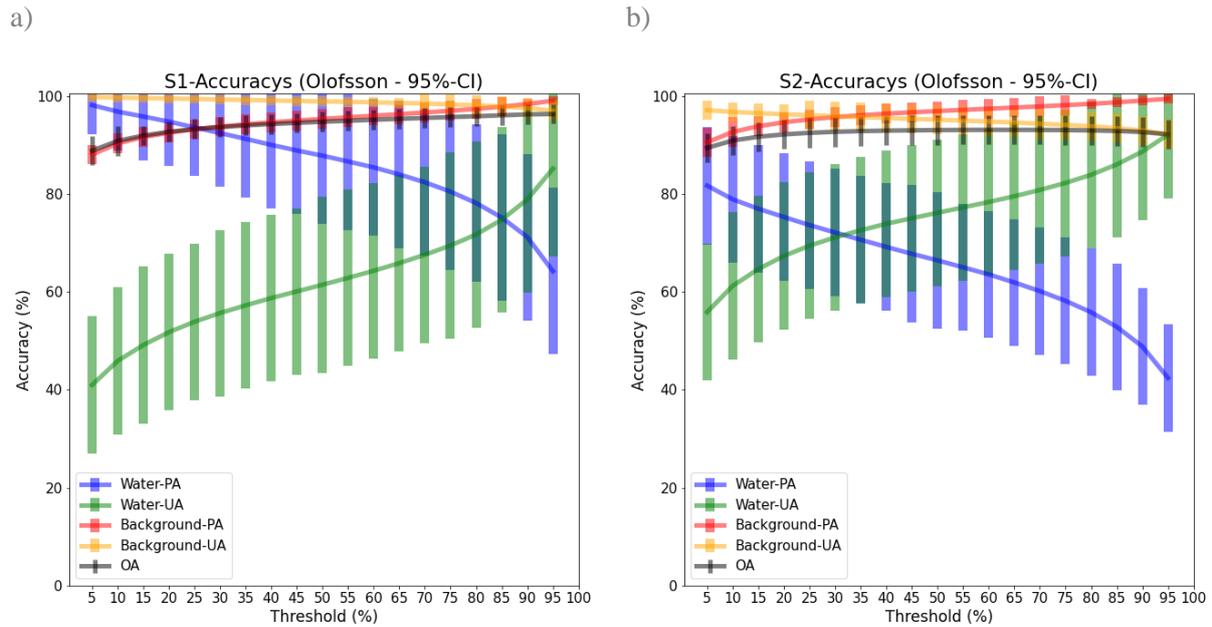

Figure 13: UNet class accuracies, user accuracy (UA), producer accuracy (PA) and overall accuracy (OA) for water and background with uncertainties, using a calculated 95% confidence interval referring to Olofsson et al. 2014. (a) UNet - Sentinel-1 validation –. (b) UNet - Sentinel-2 validation.

### 4.2.2 Confusion Matrix based Accuracy Assessment

The probabilistic model-based inundation predictions needed to be binarized using a threshold, as a binary map comparison is necessary to obtain the confusion matrix. As an optimal value for the confidence-based threshold is unknown *a priori*, multiple thresholds are evaluated through confusion matrices to evaluate the prediction quality at different certainty levels.

Figure 14 presents the results of the comparison of the model predictions with the validation set based on the Sentinel-1 water mask. The OA is mostly aligned with the background class accuracies, which is an expected effect of the class imbalance, but the user's and producer's accuracy of the flood class allow a more robust assessment of the model performance (Figure 13).

#### 4.2.2.1 *UNet*

With a low threshold of 5%, the UNet model can predict almost the entire inundation with a PA of 98.17%, while maintaining a reasonable UA of 40.98% for the flood class. Increasing the threshold, i.e., only pixels with a very high classification certainty are defined as water, the UA rises to a maximum of 85.18%, while still being able to detect 64.24% of the water. Even though the model predictions do not achieve a UA value close to 100% using high certainty thresholds, the results indicate that the CNN makes full use of the probabilistic segmentation model, as the UA scales proportionally to the selected threshold (Figure 14a). Yet, the error bars derived through methods outlined in Olofsson et al. (2014), indicate a high uncertainty within the evaluation of the water class, likely due to the class imbalance, due to the smaller area proportion of water compared to the background. This hypothesis is supported by the low uncertainties observed in the background UA/PA values, due to an abundance of samples of this class. For instance, the 95% confidence intervals (CI) for the PA for the water class range from 6 to 17%, while for the background class the range between 1 to 2%. Similarly for the UA, the CI for the water class lies between 14 to 18% while for the background it ranges from 1 to 2%. Higher thresholds seem to increase the evaluation uncertainty within the water class and decrease uncertainty in the background class, while for lower thresholds the opposite phenomenon can be observed. Figure 15 provides a reasonable explanation for this, as the decrease in the proportion of the water class can be observed with increasing thresholds. As the size of the water class decreases with higher thresholds, the proportion of the background obviously increases, thus, increasing the uncertainty in the evaluation of the foreground through the confusion matrix and leading to larger CIs (D. M. Chen & Wei, 2009).

For a completely independent validation, the results are also compared with the Sentinel-2 water mask, which was not used at all during the training process. Additionally, optical or multispectral data shows flooding differently than SAR and is often seen as a benchmark evaluation data when cloud-free images are available (G. Schumann et al., 2018). The evaluations of the model predictions against the Sentinel-2 water mask (Figure 13b), indicates a lower PA and a higher UA on an average across all thresholds. This is probably caused by the nature of the difference between the different ground truth masks, as the Sentinel-1 SAR image shows much less inundation compared to Sentinel-2. This implies that since the models were trained with the Sentinel-1, some areas which are inundated according to Sentinel-2 are predicted by the model to be dry. The higher UA indicates that many pixels which are considered to be falsely classified as water (FP) when comparing against the Sentinel-1 image are actually considered to be flooded according to the Sentinel-2 image. As the overall proportion of the water class is higher in the Sentinel-2 image, the CIs also improve in this assessment indicating higher confidence in the accuracy assessment.

a)                                                                     b)

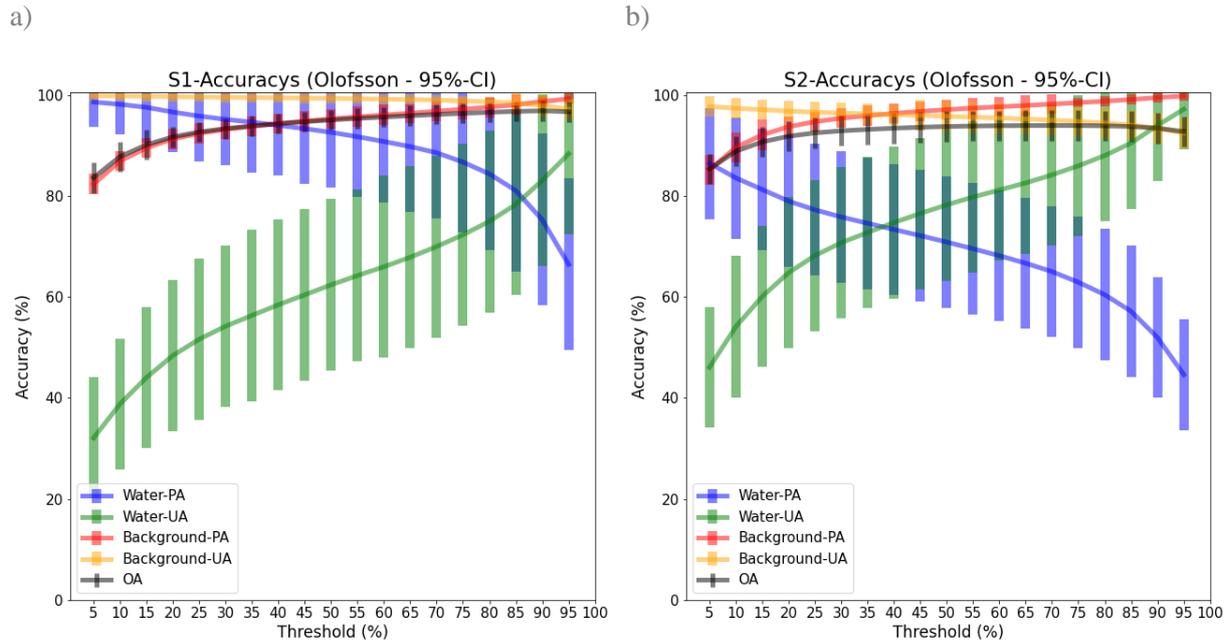

Figure 14: SegNet class accuracies, user accuracy (UA), producer accuracy (PA) and overall accuracy (OA) for water and background with uncertainties, using a calculated 95% confidence interval referring to Olofsson et al. 2014. (a) SegNet - Sentinel-1 validation. (b) SegNet - Sentinel-2 validation.

### 4.2.2.2 SegNet

Only slight differences can be observed between the UNet and SegNet predictions for the validation dataset when examining the threshold-based binary maps (Figure 14). The OA and the PA for the background are similar as expected due to the large class imbalance, which also leads to much higher PA and UA values for the background as compared to inundation. An evaluation against the Sentinel-1 based water mask, yields similar PA values for the water class of the SegNet and UNet predictions at various thresholds. However, this comparison for the UA reveals notably lower values for the water class at lower thresholds (starting at 32.09%, ~9% lower than the UNet). Conversely, for higher thresholds the PA and UA of water are both slightly higher compared to UNet, but the differences remain within the range of the (rather large) confidence intervals for the water class accuracies.

Validating against the Sentinel-2 based flood mask, shows similar effects in terms of a higher UA but a lower PA for the water class, as compared to the metric values obtained for the evaluation with the Sentinel-1 water mask (Figure 14). Even though the training evaluation showed a much more stable training rate for the UNet with a less fluctuations in the testing loss versus the SegNet (Section 4.1), the SegNet still achieves metric values similar to those of the UNet. One of the reasons for could be the low sensitivity of these binary pattern matching metrics towards the change in the inundated area between the different maps (See Dasgupta & Waske, 2022). At higher thresholds, the SegNet outperforms the UNet slightly, and vice versa at lower thresholds.

## 4.3 Feature Importance Analysis

The importance of different input features which were used to train the CNNs, in determining the output predictions of the models were examined using the PFI method (see Section 3.2), where the manipulated datasets were used as inputs to the pre-trained classifiers. Results were evaluated based on the PR-AUC calculated using the validation set derived from the Sentinel-1 based water mask (Figure 6b). Feature importance is ranked by the absolute difference between the PR-AUC obtained using the manipulated data as the input versus that resulting from the use of the original dataset (Table 2).

Table 2: (a) UNet permutation feature importance ranking. (b) SegNet permutation feature importance ranking.

a)

| Manipulated Layer | AUC-PRC | AUC Difference |
|---|---|---|
| Sentinel 2-NIR | 0,07 | -0,78 |
| Sentinel 2-Red | 0,09 | -0,77 |
| Sentinel 2-RedEdge4 | 0,16 | -0,70 |
| Sentinel 2-RedEdge3 | 0,18 | -0,67 |
| Sentinel 2-Blue | 0,20 | -0,65 |
| Sentinel 2-Green | 0,24 | -0,61 |
| Sentinel 2-RedEdge2 | 0,24 | -0,61 |
| Sentinel 2-SWIR1 | 0,24 | -0,61 |
| Sentinel 2-RedEdge1 | 0,33 | -0,53 |
| COP30-DEM-Elevation | 0,34 | -0,52 |
| Sentinel 2-SWIR2 | 0,54 | -0,32 |
| COP30-DEM-Slope | 0,59 | -0,26 |
| ESA-LULC | 0,80 | -0,06 |
| Sentinel 1-VV | 0,84 | -0,02 |
| GPM-IR-Precipitation | 0,84 | -0,01 |
| GPM-CAL-randomError | 0,86 | 0,00 |
| Sentinel 1-VH | 0,85 | 0,00 |
| GPM-CAL-Precipitation | 0,86 | 0,00 |
| COP-Soil-Moisture | 0,85 | 0,00 |
| COP-Soil-Moisture-Uncertainty | 0,85 | 0,00 |
| GPM-HQ-Precipitation | 0,85 | 0,00 |

b)

| Manipulated Layer | AUC-PRC | AUC Difference |
|---|---|---|
| Sentinel 2-Red | 0,08 | -0,78 |
| Sentinel 2-SWIR2 | 0,08 | -0,78 |
| Sentinel 2-RedEdge2 | 0,11 | -0,75 |
| Sentinel 2-NIR | 0,11 | -0,75 |
| Sentinel 2-RedEdge4 | 0,13 | -0,73 |
| Sentinel 2-RedEdge3 | 0,17 | -0,69 |
| Sentinel 2-Blue | 0,20 | -0,66 |
| Sentinel 2-Green | 0,24 | -0,63 |
| COP30-DEM-Elevation | 0,39 | -0,47 |
| Sentinel 2-SWIR1 | 0,42 | -0,44 |
| COP30-DEM-Slope | 0,45 | -0,41 |
| Sentinel 2-RedEdge1 | 0,60 | -0,26 |
| ESA-LULC | 0,80 | -0,06 |
| GPM-IR-Precipitation | 0,84 | -0,02 |
| GPM-CAL-randomError | 0,86 | 0,01 |
| GPM-HQ-Precipitation | 0,86 | 0,00 |
| Sentinel 1-VH | 0,86 | 0,00 |
| COP-Soil-Moisture | 0,86 | 0,00 |
| COP-Soil-Moisture-Uncertainty | 0,86 | 0,00 |
| GPM-CAL-Precipitation | 0,86 | 0,00 |
| Sentinel 1-VV | 0,86 | 0,00 |

### 4.3.1 Feature Importance Rankings

The pre-floods Sentinel-2 bands and the DEM emerge as the most important input features for both the CNN models. The Sentinel-1 and LULC data appear to be less important, as does the precipitation evident from the low change in the PR-AUC values when replacing these bands with noise. Interestingly, there is no change in accuracy when exchanging the soil moisture data with noise, meaning that soil-moisture has no influence at all on the model predictions. As soil moisture and precipitation upstream are expected to provide information on the water entering the domain, this is a rather unsurprising outcome, since only the downstream soil moisture and precipitation could be included here. Even though the downstream precipitation and soil moisture could partly influence the inundation, the importance of these variables for the model could have been further reduced due to the large resolution mismatch between these inputs and the output. Previous studies have shown the correlation of soil moisture and flood inundation (e.g. Du et al. 2021), and indeed, the hydrological relevance of saturated soils for inundation is rather obvious. However, the very coarse spatial resolution of the soil moisture data resulting in low variability within the prediction patches, as well as the lack of upstream hydrological variables, is expected to have caused this independence between the input and the predictions. In future, ways to incorporate the rainfall and antecedent catchment conditions upstream should be prioritized.

Figure 17 shows the results obtained when manipulating all (image) layers of soil moisture or precipitation with the pre-trained UNet in (A) and SegNet in (B). For both models, the predictions obtained using manipulated soil moisture layers are identical to the predictions using the original data, implying that neither of the models use any trained weights influenced by soil moisture to predict the flood inundation. Precipitation does influence the results slightly, which is well aligned with the ranking based on the PR-AUC difference (Table 2). This feature importance ranking could be used to reduce the number of input features with low importance, such as the pre-flood Sentinel-1 SAR image or GPM layers, to reduce the computational requirements and optimize training over larger areas. Future studies should also compute Relative Feature Importance to further identify and remove redundancies from the input feature datasets, allowing the design of more streamlined and computationally efficient prediction models.

A)                                                    B)

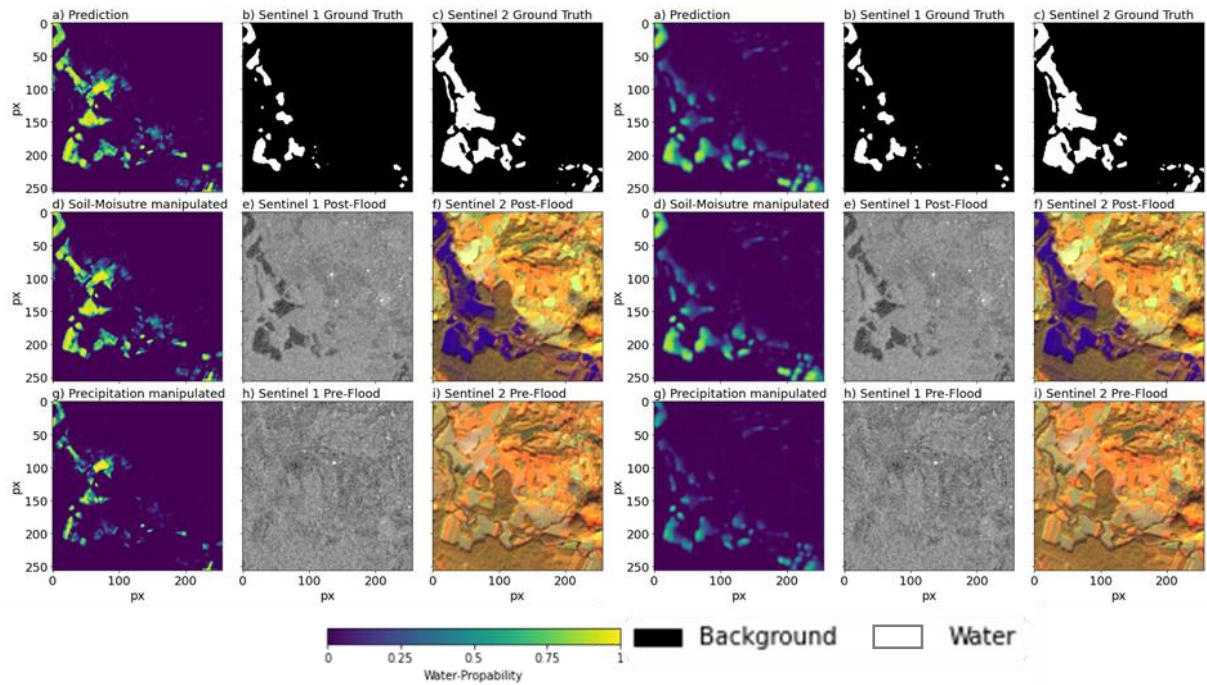

Figure 15: (A) UNet permutation feature importance example by exchanging all Soil-Moisture or all GPM Bands with random noise. (B) SegNet permutation feature importance example by exchanging all soil-moisture or all GPM layers with random noise. (a) Prediction. (b) Sentinel-1 ground truth. (c) Sentinel-2 ground truth. (d) UNet prediction without Soil-Moisture but random noise instead. (e) Sentinel-1 post-flood image in VV polarization. (f) Sentinel-2 post-flood image as false-color-composite using (R: RedEdge4, G: SWIR1, B: Red). (g) UNet prediction without GPM but random noise instead. (h) Sentinel-1 pre-flood image in VV polarization. (i) Sentinel-2 pre-flood image as false-color-composite using (R: RedEdge4, G: SWIR1, B: Red).

### 4.3.2 Influencing Factors and Perspectives

The validation showed that both trained CNNs are able to map inundated areas at 10m spatial resolution, achieving high accuracies with a PR-AUC of ~0.85 using the given input data, consisting of optical and SAR pre-flood images, DEM, LULC, precipitation and soil moisture. Only the last two inputs, i.e. precipitation and soil moisture, actually contain information about the water coming into the domain. However, these have the least importance for the prediction, according to the results of the PFI ranking, meaning that the inundation was mapped without using this flood related data. One possible reason for this is the small variation within the coarse resolution soil moisture and rainfall data. However, since only a PFI is performed to evaluate the feature importance, it is also possible that the soil moisture and precipitation compensate for one another when replacing one of the datasets with noise. In order to further investigate, the utility of satellite soil moisture datasets for high-resolution flood mapping, more variant data must be acquired for training the models and the relative feature importance must also be considered. While some high-resolution soil moisture data already exists, e.g. the SMAP-Sentinel-1 L2 Soil moisture product at 3km resolution (Das et al., 2017), these are limited by the Sentinel-1 revisit time and thus, are unable to provide daily scale data as was required for this proof-of-concept study. Here it was assumed that previous rainfall is included within the current soil moisture, and thus, only the precipitation data for the day of prediction is used as an input. As the precipitation data have a much higher resolution (0.1°) as compared to soil moisture (0.25°), it could be worth retraining the classifier with a previous precipitation time-series.

The most important features were the optical Sentinel-2 pre-flood image and the DEM (Table 2). The Sentinel-2 pre-flood image is assumed to be an information source about land cover, i.e. in hydrological terms the surface roughness which characterizes the resistance to flow. Thus, the feature importance rankings suggest that land cover and elevation are decisive factors for inundation, which is well-aligned with the hydrological principles of flood routing. Given the high importance of the pre-flood Sentinel-2 image and DEM (Table 2), it seems that these two data sources can directly be used to predict flood inundation given a target, i.e. an inundation map from SAR or optical post-flood images. This could be particularly useful to "look under the clouds" in Sentinel-2 images, where theoretically a DEM could be used to predict the hidden inundation. Finally, the CNNs were trained and tested only on one flood event in this study, where they were able to capture inundation patterns with high levels of detail. However, the transferability of the model and generalizability to other climate or dominant land cover zones needs to be tested. Future work will also evaluate the predictive performance of the CNNs for different flood stages, i.e. rising limb, peak, or receding limb of the hydrograph, to better understand the influencing factors for deep learning based inundation predictions.

## 5   Conclusions and Outlook

The main objective of our study was to assess if multi-sensor multi-resolution remote sensing data enable flood inundation predictions at the S1 spatial resolution in the absence of a corresponding S1 image. This is confirmed by the results and the possibility of predicting daily scale high-resolution inundation using temporally frequent publicly available global datasets and deep learning is demonstrated for the first time. Another objective of the study was to assess the agreement between simulated and observed inundation during the event, where accuracies over ~0.85 PR-AUC were attained for validation against independent high-resolution cloud-free optical imagery from S2. A convolutional neural network was trained for this task using high resolution SAR based flood maps, but without any high resolution optical or microwave data showing the flood inundation used as inputs, such that the prediction time-scale could be independent of the revisit of these satellites. Leveraging multi-sensor remote sensing data including pre-flood images from Copernicus Sentinel-1 and 2 satellites, (relatively) static ancillary data such land use and elevation, as well as dynamic satellite hydrology datasets such as soil moisture and precipitation, both CNN models were able to predict high resolution inundation preserving high- and low-level spatial detail for a variety of complex situations. Predictions were evaluated against water masks derived from the Sentinel-1 and Sentinel-2 post-flood images, with a high overall value ~0.85 achieved for the area under the precision-recall curve (PR-AUC). Future work should further examine the generalizability and transferability of such models by application to multiple diverse study sites across the globe. Furthermore, the impact of flood stage (early, peak, or late) and magnitude on the predictive capacity of the models should also be investigated.

The experiments presented herein demonstrate that deep learning based models can successfully bridge the gap between Synthetic Aperture Radar acquisitions and the daily observation frequency required for observing flood inundation dynamics. Both CNN models were able to capture fine scale details of the inundation, implying that deep learning combined with multi-sensor remote sensing could support daily cadence in observing floods with significant implications for flash flooding and parametric insurance products.